\documentclass[12pt]{article}
\pdfoutput=1
\usepackage{dcolumn}% Align table columns on decimal point
\usepackage{bm}% bold math
\usepackage{graphicx}% Include figure files
\usepackage{amssymb,amsmath}
\usepackage{multirow}
\usepackage{afterpage}
\usepackage{cite,color,url}
\usepackage[colorlinks=true
,urlcolor=magenta
,anchorcolor=blue
,citecolor=blue
,filecolor=blue
,linkcolor=blue
,menucolor=blue
,linktocpage=true
,pdfproducer=medialab
,pdfa=true
]{hyperref}
\usepackage{cleveref}
\usepackage{tikz-feynman}
\usetikzlibrary{trees}
\usetikzlibrary{decorations.pathmorphing}
\usetikzlibrary{decorations.markings}

\usepackage{array,booktabs}
\usepackage{subcaption}

\usepackage{slashed}
\usepackage{epsfig,psfrag,rotating,soul}
\usepackage{rotfloat}
\usepackage[font={small}]{caption}
\usepackage{colortbl}
\usepackage{array,booktabs}
\usepackage{subcaption}
\usepackage{contour}
\usepackage[normalem]{ulem}

\evensidemargin \oddsidemargin
\allowdisplaybreaks

\let\OLDthebibliography\thebibliography
\renewcommand\thebibliography[1]{
  \OLDthebibliography{#1}
  \setlength{\parskip}{0pt}
  \setlength{\itemsep}{0pt plus 0.3ex}
}

\newcommand{\beq}{\begin{equation}}
\newcommand{\eeq}{\end{equation}}

\usepackage{xspace}

\newcommand{\ba}{\begin{array}}
	\newcommand{\ea}{\end{array}}
\newcommand{\bd}{\begin{displaymath}}
	\newcommand{\ed}{\end{displaymath}}
\newcommand{\besub}{\begin{subequations}}
	\newcommand{\eesub}{\end{subequations}}
\newcommand{\bea}{\begin{eqnarray}}
	\newcommand{\eea}{\end{eqnarray}}

\def\g{\gamma}

\def\q2 {q^2}

\def\bt{\begin{table}}
\def\et{\end{table}}

\definecolor{mygray}{gray}{0.85} 
\definecolor{myblue}{cmyk}{0.65, 0.37, 0.0, 0.19}

\begin{document}
\thispagestyle{empty}

\def\thefootnote{\fnsymbol{footnote}}

\vspace*{1cm}

\begin{center}

\begin{Large}
\textbf{\textsc{Probing ALP-portal Fermionic Dark Matter at\\ the $e^+e^-$ Colliders}}
\end{Large}

\vspace{0.5cm}

{Subhaditya ~Bhattacharya$^1$%
\footnote{{\tt \href{mailto:subhab@iitg.ac.in}{subhab@iitg.ac.in}}}%\sc
, Sahabub ~Jahedi$^{2,3}$%
\footnote{{\tt \href{mailto:sahabub@m.scnu.edu.cn}{sahabub@m.scnu.edu.cn}} (Corresponding Author)}%
, Soumen Kumar Manna$^{1,4}$%
\footnote{{\tt \href{mailto:skmanna2021@gmail.com}{skmanna2021@gmail.com}}}%
, Arunansu Sil$^1$%
\footnote{{\tt \href{asil@iitg.ac.in}{asil@iitg.ac.in}}}%
}

\vspace*{.5cm}

{\it
		
	$^1$Department of Physics, Indian Institute of Technology Guwahati, Assam 781039, India
	
	$^2$State Key Laboratory of Nuclear Physics and Technology, Institute of Quantum Matter, South China Normal 
	University, Guangzhou 510006, China\\
	
	$^3$Guangdong Basic Research Center of Excellence for Structure and Fundamental Interactions of Matter, Guangdong 
	Provincial Key Laboratory of Nuclear Science, Guangzhou 510006, China\\
	
	$^4$Department of Physics, Indian Institute of Technology Bombay, Mumbai 400076, India
	
}

\end{center}

\vspace{0.1cm}

\begin{abstract}
Axion-like particles (ALPs) are promising candidates for mediating interactions between a dark sector and the Standard Model (SM). 
In this work, considering the effective interactions of ALPs with the SM gauge bosons and a fermion dark matter (DM), 
we explore the DM relic satisfied parameter space and assess its testability through indirect searches. The effect of early kinetic decoupling is also discussed in the resonant regime. The potential of probing such ALP-portal fermionic DM at the electron-positron colliders is investigated with the mono-photon plus missing energy final states. We show that a spectacular distinction between the signal and SM background is possible via the missing energy variable, the seed of which lies in the  ALP-photon interaction, which also governs the relic density of DM. We further discuss the sensitivity of ALP-photon coupling using the $\chi^2$ analysis at the future electron-positron collider specifications.

\end{abstract}
%%%%%%%%%%%%%%%%%%%%%%%%%%%

\def\thefootnote{\arabic{footnote}}
\setcounter{page}{0}
\setcounter{footnote}{0}

\newpage

%%%%%%%%%%%%%%%%%%%%%%%%%%%%
\section{Introduction}
\label{sec:intro}
%%%%%%%%%%%%%%%%%%%%%%%%%%%%

Axion like particles (ALPs) are known to be a well-explored example of pseudo-Nambu-Goldstone bosons (pNGB), commonly arising within theories featuring spontaneously broken global symmetry. Originally introduced as QCD axion, the pNGB of global axial Peccei-Quinn {symmetry $U(1)_{\rm PQ}$ was} proposed to address the strong CP problem \cite{Peccei:1977hh,Peccei:1977ur,Weinberg:1977ma,Wilczek:1977pj}, where the symmetry encounters explicit breaking by the QCD non-perturbative effects at the QCD confinement scale $\Lambda_{\rm QCD}\approx 150~ {\rm MeV}$. ALPs, on the other hand, represent a broader category of pNGBs, associated to a similar PQ-like global $U(1)$ symmetry with its potential generated by some QCD-like (exotic) strongly interacting hidden sector and hence not primarily {to solve the strong CP problem. Still, their significance lies in numerous extensions of the Standard model (SM), $e.g.$, string theory \cite{Svrcek:2006yi,Arvanitaki:2009fg,Demirtas:2018akl}. Being pNGBs in nature, ALPs are expected to be very light and couple to the SM gauge bosons and/or fermions at an effective level, suppressed by the spontaneous symmetry breaking scale $(f_a)$ of the global symmetry (or, the ALP decay constant). As ALPs' masses may stem from non-perturbative effects of the hidden sector or from explicit symmetry breaking, {their masses and $f_a$ (hence, couplings to the SM particles)} are independent to each other and span across several orders of magnitude, unlike QCD axion. {It is known that in certain regions} of parameter space (predominantly below MeV scale), ALPs can act as cold dark matter (CDM), originating from non-thermal production through coherent oscillations in the early universe, commonly known as misalignment mechanism \cite{Preskill:1982cy,Abbott:1982af,Dine:1982ah,Turner:1983he,Arias:2012az}. Alternately, ALPs can {also serve} as a portal to the dark sector, as discussed in refs. \cite{Dolan:2014ska,Dolan:2017osp,Hochberg:2018rjs,CidVidal:2018blh,Bharucha:2022lty,Ghosh:2023tyz,Fitzpatrick:2023xks,Dror:2023fyd,Acanfora:2023gzr,Armando:2023zwz,Allen:2024ndv}.

Over several decades, extensive searches and observations have been carried out across diverse mass and coupling (phenomenologically, most important one is ALP-photon coupling) ranges of axion and ALPs, leading to stringent constraints on the parameter space. For example, very light long-lived ALPs $(m_a\lesssim\mathcal{O}(10)~{\rm eV})$ are mostly constrained by the light-shining-through-a-wall (LSW) experiments \cite{Redondo:2010dp} and other Helioscope experiments like SUMICO \cite{Inoue:2008zp, Graham:2015ouw} and CAST \cite{CAST:2017uph}. Furthermore, in the intermediate ALP mass range (keV-MeV), a plethora of cosmological constraints e.g. alterations in primordial big bang nucleosynthesis (BBN) and number of relativistic degrees of freedom (d.o.f) \cite{Masso:1995tw,Cadamuro:2011fd,Poulin:2015opa,Hufnagel:2018bjp, Depta:2020wmr,Balazs:2022tjl}, distortions of Cosmic Microwave background (CMB) \cite{Hu:1993gc,Lucca:2019rxf} and fluctuations in extragalactic background light (EBL) \cite{Overduin:2004sz} put constraints, particularly for small ALP-photon couplings. Above the MeV scale, ALPs with significant ALP-photon couplings have attracted experimental interest. In this regime, ALPs are sensitive to be probed in various flavor and collider experiments like BaBar \cite{BaBar:2010eww}, LEP \cite{Jaeckel:2015jla} and LHC \cite{Mimasu:2014nea} as well as in current and future beam-dump experiments \cite{Riordan:1987aw,Bjorken:1988as,Dobrich:2015jyk}.

In this work, we are primarily interested in the phenomenology of ALPs with mass above GeV scale and its search at the high energy collider frontiers. Notably, GeV scale ALP cannot serve as a viable DM candidate as its decay into two photons leads to instability over cosmological timescale. Instead, we explore its role as mediator between DM and SM sector, where a Dirac fermion ($\Psi$) serves as the DM candidate. Considering a scenario where the GeV scale ALPs only interact with electroweak gauge bosons, DM can achieve the correct relic abundance through annihilation into two photons during freeze-out. However, stringent constraints from indirect detection experiments can significantly restrict the parameter space. These constraints can be evaded if DM annihilation is resonantly enhanced during freeze-out, allowing the correct relic abundance to be achieved for $f_a\sim \mathcal{O}(200)$ GeV. Interestingly, such GeV-scale ALPs with $f_a\sim \mathcal{O}(200)$ GeV can also be probed at the future lepton colliders.
Once produced, ALPs can decay into photons, charged leptons, light hadrons, or jets, leading to multi-photon \cite{Bauer:2018uxu,Zhang:2021sio,Lu:2022zbe}, multi-lepton \cite{Yue:2021iiu,Calibbi:2022izs,Yue:2023mjm}, and multi-jet \cite{Yue:2022ash} signals that motivate collider searches. Considering the ALP as a long-lived particle, the study of the mono-photon signal has been carried out in \cite{Bao:2025tqs}. An intriguing possibility in our setup is that ALP can dominantly decay into two DM particles (considering $m_a\gtrsim 2 m_{\Psi}$), which would show up as missing energy ($E_{\rm miss}$) or 
missing transverse momentum $\slashed{E}_T$ at colliders. The signal cross-section being proportional to the DM-ALP coupling, which also contributes to the DM relic density, observation of such a DM signal can indicate towards the cosmological imprint of the DM.

Our proposed signal for probing this model at future electron-positron collider like ILC \cite{Behnke:2013xla,ILCInternationalDevelopmentTeam:2022izu} is mono-photon ($\gamma$) plus $E_{\rm miss}$, where we look for an associated production of ALP with photon, with ALP predominantly decaying to DM. Thanks to the absence of the QCD background, PDF uncertainty, and the possibility of beam polarisation, lepton colliders allow us to investigate different new physics (NP) scenarios in a pristine environment, with less SM contamination. Further, a spectacular distinction in the $E_{\rm miss}$ distribution of this specific signal from the SM background allows us to estimate the sensitivity of ALP-SM couplings to a high degree of accuracy.

Notably, invisible particle search using mono-$\gamma$ + $E_{\rm miss}$ signature has been performed previously at LEP~\cite{DELPHI:2008uka}. The results have been recast in terms of the photon-energy observable~\cite{Fox:2011fx}, 
and subsequently reinterpreted to constrain ALP-photon coupling via invisible ALP decay~\cite{Darme:2020sjf}. Apart, BaBar experiment has also probed invisible ALP decay in $e^+e^-$ collisions through the same channel ~\cite{Izaguirre:2016dfi}. At the high-intensity frontier of the BaBar experiment, invisible ALP decay to DM has been probed via $B \to K a$ \cite{BaBar:2017tiz}. Belle-II has also provided a sensitivity limit on ALP-photon coupling for invisible ALP decays considering mono-$\gamma$ + $E_{\rm miss}$ channel~\cite{Dolan:2017osp,Acanfora:2023gzr}. However, ALP-photon coupling has never been probed in the context of ALP-portal DM model at the future high energy $e^+e^-$ colliders. Thanks to a clear signal, we show that ALP-photon coupling can be probed upto a precision of $\left(\sim \mathcal{O}(10^{-5})\right)$ GeV$^{-1}$ at the ILC, 
which is difficult to achieve at the high luminosity LHC (HL-LHC).

This paper is structured as follows: In \Cref{sec:pheno}, we outline the phenomenological framework of the study. The DM phenomenology is discussed in \Cref{sec:DM}. A detailed collider analysis is presented in \Cref{sec:col}. In \Cref{sec:coll.sen}, we briefly explore standard $\chi^2$ analysis and evaluate the collider accuracy with which the ALP-photon coupling can be estimated. We conclude our findings in \Cref{sec:con}, and in \Cref{sec:uv.comp}, we discuss a possible UV completion of the model framework.

%%%%%%%%%%%%%%%%%%%%%%%%%%%%%%%%%%%%%%%
\section{Phenomenological Framework}
\label{sec:pheno}
%%%%%%%%%%%%%%%%%%%%%%%%%%%%%%%%%%%%%%
An ALP is typically described as a pNGB field $a$, arising from the spontaneous breaking of a global axial $U(1)$ symmetry via a complex scalar field $\Phi~=(f_a+\rho) e^{ia/f_a}/{\sqrt{2}}$, where $f_a$ is the symmetry breaking scale. The leading interactions of such pseudoscalar ALP with the SM particles can be described by dimension-5 effective operators at low energy. In this work, we consider the following ALP Lagrangian \cite{Brivio:2017ije,Bauer:2017ris,Allen:2024ndv}:
\beq
\mathcal{L}_{\rm eff}= \mathcal{L}_{\rm ALP} +\mathcal{L}_{\rm ALP-SM}+\mathcal{L}_{\rm ALP-DM},
\label{eq:ALP-eff}
\eeq
where 
\bea
\mathcal{L}_{\rm ALP}&=&\frac{1}{2}\partial_\mu a \partial^\mu a -\frac{1}{2} m_a^2 a^2,
\label{eq:L-ALP}\\
\mathcal{L}_{\rm ALP-SM}&=&-\frac{C_{aB}}{4} a B_{\mu\nu}\tilde{B}^{\mu\nu}-\frac{C_{aW}}{4} a W_{\mu\nu}\tilde{W}^{\mu\nu}.
\label{eq:ALP-SM}
\eea
Here $B_{\mu \nu}$ and $W_{\mu \nu}$ are field strength tensors corresponding to $U(1)_Y$ and $SU(2)_L$ gauge groups respectively, while the dual field strength tensors are defined as $\tilde{B}^{\mu\nu}=\frac{1}{2}\epsilon^{\mu\nu \alpha \beta} B_{\alpha\beta}$ and so on. The coupling constants, $C_{ai}~(\propto {\alpha_i}/{\pi f_a})$ for $i=B,W$, are the Wilson coefficients.
The ALP mass ($m_a$) in Eq. \eqref{eq:L-ALP} is a free parameter, which can arise from either non-perturbative dynamics or explicit breaking of the underlying symmetry. As evident from Eq. \eqref{eq:ALP-SM}, we primarily focus on the ALP interaction with SM electroweak gauge bosons, while neglecting any direct couplings to SM fermions and gluons. This setup can be naturally realized, analogous to the KSVZ axion model \cite{Kim:1979if,Shifman:1979if}, which {is elaborated} in appendix \ref{sec:uv.comp}. The absence of ALP-fermion couplings also prevents flavor-changing processes at the tree level\footnote{However, at the loop-level, flavor violation can be induced via the SM $W$-boson interactions, leading to constraints from meson decays, see e.g. Ref \cite{Izaguirre:2016dfi}.}, which are strongly constrained by rare decay searches \cite{Fitzpatrick:2023xks,Acanfora:2023gzr,Dolan:2017osp,Allen:2024ndv}.

After electroweak symmetry breaking, the Lagrangian that governs the ALP gauge boson couplings can be described (following Eq. \eqref{eq:ALP-SM}) as 
\beq
\mathcal{L}_{\rm ALP-SM}=-\frac{g_{a\gamma\gamma}}{4} a F_{\mu\nu}\tilde{F}^{\mu\nu}-\frac{g_{aZZ}}{4} a Z_{\mu\nu}\tilde{Z}^{\mu\nu}-\frac{g_{aWW}}{4} a W^+_{\mu\nu}\tilde{W}^{-\mu\nu}-\frac{g_{a\gamma Z}}{4} a F_{\mu\nu}\tilde{Z}^{\mu\nu},
\label{eq:ALP-SM2}	
\eeq
where $F_{\mu\nu},Z_{\mu\nu} $ and $W_{\mu\nu}$ are corresponding to the field strength tensors of photon, $Z$ and $W$ bosons respectively, and the effective couplings involved in Eq.~\eqref{eq:ALP-SM2} follow the relations
\bea
g_{a\gamma\gamma}&=& C_{aB}~ {\rm cos}^2\theta_W + C_{aW}~ {\rm sin}^2
\theta_W,
\label{eq:ga}\\
g_{aZZ}&=& C_{aB} ~{\rm sin}^2\theta_W + C_{aW}~ {\rm cos}^2
\theta_W,\\
g_{a\gamma Z}&=& (C_{aW}-C_{aB})~ {\rm cos}\theta_W ~{\rm sin}
\theta_W, \\
g_{aWW}&=& C_{aW},
\eea
with $\theta_W$ being the weak mixing angle. For the economy of our calculation, we assume $C_{aB}=C_{aW}$ 
\footnote{We do not fix their individual values, apart from their parametric dependence $\propto {\alpha_i}/{\pi f_a}$ for $i=B,W$, as mentioned earlier. The precise values depend on anomaly coefficients that can arise from integrating out heavy fermions charged under the $U(1)$ symmetry as well as the electroweak gauge groups (see Appendix~\ref{sec:uv.comp} for a detailed discussion).}. 
We parametrize the interaction in terms of the ALP-photon physical coupling $g_{a\gamma\gamma}$, which, after the electroweak symmetry breaking, takes the form $g_{a\gamma\gamma}={\alpha_{\rm EM}}/{(\pi f_a)}$. The assumption $(C_{aB}=C_{aW})$ then greatly simplifies our analysis with $g_{a\gamma\gamma}=g_{aZZ}=g_{aWW}$ and $g_{a\gamma Z}=0$. We may note further that for the mass hierarchy of our interest $(m_a<M_Z)$, $C_{aB}\neq C_{aW}$ does not significantly affect the DM phenomenology. However, such a relaxation opens up an additional decay mode of the $Z$ boson, $Z \to \gamma a$, which is constrained by the LEP 
data with an upper bound on the branching ratio $\text{BR}(Z \to \gamma a) < 1.1 \times 10^{-6}$~\cite{L3:1997exg,Dolan:2017osp}. This inequality also modifies the constraint derived from Fermi-LAT observations. From the experimental perspective, the ALP interaction with two photons is particularly significant {for $m_a<M_W,M_Z$ (with $M_W$ and $M_Z$ being the masses of $W$ and $Z$ bosons, respectively)}, resulting the decay width of ALP as \cite{Allen:2024ndv}
\beq
\Gamma_a=\frac{g_{a\gamma\gamma}^2 m_a^3}{64\pi},~~~{\rm with}~~g_{a\gamma\gamma}=\frac{\alpha_{\rm EM}}{\pi f_a},
\label{eq:gagg}
\eeq
where $\alpha_{\rm EM}$ denotes the fine-structure constant, arising from the $U(1)_Y$ and $SU(2)_L$ gauge field strength operators after EWSB, following Eq.~\eqref{eq:ga}.

We now focus our attention to ALP-DM interaction. As stated earlier, we consider a Dirac fermion $\Psi$, charged under the same global $U(1)$ symmetry, to serve as the DM candidate. The DM mass term and its shift-symmetric interaction with ALPs can then be expressed as 
\beq
\mathcal{L}_{\rm ALP-DM}=-m_\Psi \bar{\Psi} \Psi -\frac{c_\Psi}{2f_a} (\partial_\mu a) \bar{\Psi} \gamma^\mu\gamma_5 \Psi,
\label{eq:ALP-DM}
\eeq
where $c_\Psi$ is a dimensionless coefficient. The origin of such ALP-DM interaction,  {after the spontaneous breaking of $U(1)$}, is discussed in appendix \ref{sec:uv.comp}. In this setup, a residual $Z_2$ after $U(1)$ breaking is present which naturally stabilizes the DM\footnote{A Majorana fermion can be another choice of such fermion DM candidate, which can be stabilized by ${Z}_2$ symmetry \cite{Acanfora:2023gzr,Dolan:2017osp,Ghosh:2023tyz}.}, ensuring that it is always produced in $\bar{\Psi}\Psi$ pairs \cite{Bharucha:2022lty}. Choosing specific mass hierarchies as: $m_a\gtrsim 2 m_{\Psi}$ and $m_a<M_W, M_Z$, in the following sections, we explore the phenomenology of ALP-mediated DM interactions and their potential detection in collider searches.

%%%%%%%%%%%%%%%%%%%%%%%%%%%%%%%%%%%%%%%%%%
\section{Dark Matter Phenomenology}
\label{sec:DM}
%%%%%%%%%%%%%%%%%%%%%%%%%%%%%%%%%%%%%%%%%

In this section, we investigate the framework discussed above, mainly emphasizing on the possibility of DM thermal freeze out. First, following Ref. \cite{Allen:2024ndv}, we analyze the standard picture of DM freeze-out considering a benchmark set of parameters. Furthermore, we briefly discuss the effect of early kinetic decoupling (EKD) on DM relic abundance relative to the benchmark values.

%%%%%%%%%%%%%%%%%%%%%%%%%%%%%%%%%%%%%%%%%%%%%%%%%%%%%%%%%%%%%%%%%%
\subsection{Dark matter freeze-out near resonance: standard case} 
%%%%%%%%%%%%%%%%%%%%%%%%%%%%%%%%%%%%%%%%%%%%%%%%%%%%%%%%%%%%%%%%%%

For reasonably low $f_a$, the dark sector is expected to maintain thermal equilibrium with the SM sector {while allowing DM to freeze-out} through $\Psi$ annihilations into SM particles (except SM fermions). Following Eqs. \eqref{eq:ALP-SM2} and \eqref{eq:ALP-DM}, it is evident that DM interacts with SM electroweak gauge bosons via ALP mediation. 
The DM mass being smaller than the SM gauges bosons (as specified by the chosen mass hierarchies), the dominant annihilation channel of DM freeze out turns out to be 
$\bar{\Psi}\Psi\to\g\g$ as in Fig.~\ref{DM-anni}, while other possible annihilation channels, such as $\bar{\Psi}\Psi\to W^+W^-$, $\bar{\Psi}\Psi\to ZZ$, and $\bar{\Psi}\Psi\to aa$ remain kinematically forbidden. Such {a typical DM mass range} is motivated by constraints from both indirect detection and collider searches, which we will discuss in detail later.
\begin{figure}[h!]
\begin{center}
	\begin{tikzpicture}
	\begin{feynman}
		\vertex (a) at (0, 1.7) { $\Psi$};
		\vertex (b) at (0, -1.7) {$\bar{\Psi}$};
		\vertex (c) at (2.3, 0);
		\vertex (c1) at (4.8, 0);
		\vertex (d) at (7, 1.7) {\Large $\gamma$};
		\vertex (e) at (7, -1.7) {\Large $\gamma$};
		
		\diagram* {
			(a) -- [fermion, ultra thick, arrow size=2pt] (c) -- [fermion, ultra thick, arrow size=2pt] (b),
			(e) -- [boson, ultra thick, arrow size=2pt] (c1) -- [boson, ultra thick, arrow size=2pt] (d),
			(c) -- [scalar, ultra thick, edge label=$a$] (c1)
		};
		\vertex at (c) [blob, minimum size=0.5cm, fill=gray!50] {};
		\vertex at (c1) [blob, minimum size=0.5cm, fill=gray!50] {};
	\end{feynman}
\end{tikzpicture}
\end{center}
\caption{Feynman diagram indicating the annihilation of two Dirac fermion DM particles ($\Psi$) into two photons via {ALP ($a$) mediation, responsible for DM freeze-out}.}
\label{DM-anni}
\end{figure}

The evolution of DM can be analyzed by solving the Boltzmann equation in terms of the co-moving number density $Y_\Psi~(=n_\Psi/\mathcal{S})$ and the dimensionless variable $x=m_\Psi/T$:
\bea
\frac{dY_\Psi}{dx}&=& -\frac{\mathcal{S}\langle\sigma_{\Psi\bar{\Psi}\to\gamma\gamma} v\rangle^{\rm sub}}{x H(x)}[Y_\Psi^2-{(Y_\Psi^{\rm eq})}^2]-\frac{\langle\Gamma_a\rangle Y_a^{\rm eq}}{x H(x)}{\rm BR}(a\to \Psi\bar{\Psi})\left[\frac{Y_\Psi^2}{(Y_\Psi^{\rm eq})^2}-1 \right],
\label{eq:BE}
\eea
where $H=1.66\sqrt{g_\star^\rho(T)}\frac{T^2}{M_{\rm Pl}}$ is the Hubble expansion rate of the Universe, and $\mathcal{S}=\frac{2\pi^2}{45}g_\star^\mathcal{S}(T)T^3$ denotes the entropy density of the Universe. The equilibrium abundance of the species $i$ is given by $Y^{\rm eq}_i$, expressed as 
\beq
Y_i^{\rm eq}=\frac{45 g_i}{4\pi^4 g_\star^\mathcal{S}(T)}\left(\frac{m_i}{T} \right)^2 K_2\left(\frac{m_i}{T} \right),
\eeq
where $m_i$ and $g_i=1(2)$ (for scalar (fermion)) represent the mass and internal degrees of freedom of the species $i$. The thermally averaged cross-section for $\Psi\bar{\Psi}\to\gamma\gamma$ is denoted by $\langle\sigma_{\Psi\bar{\Psi}\to\gamma\gamma} v\rangle$, can be expressed as
\beq
\langle\sigma_{\Psi\bar{\Psi}\to\gamma\gamma} v\rangle=	\frac{1}{8m_\Psi^4 (m_\Psi/x) K_2^2(x)}
\int_{4m_\Psi^2}^{\infty}\sigma_{\Psi\bar{\Psi}\to\gamma\gamma} (s-4m_\Psi^2)\sqrt{s}K_1\left( { x\sqrt{s}}/{m_\Psi} \right)  ds,
\label{eq:svavg}
\eeq
where 
\beq
\sigma_{\Psi\bar{\Psi}\to\gamma\gamma}=\frac{g_{a\gamma\gamma}^2 m_\Psi^2}{128\pi f_a^2}\frac{1}{\sqrt{1-4m_\Psi^2/s}}\frac{s^2}{(s-m_a^2)^2+m_a^2\Gamma_a^2},
\label{eq:sigma}
\eeq
with $s$ being the center-of-mass (CM) energy and $K_1 (K_2)$ denotes the modifies Bessel's function of first (second) kind. Similarly, $\langle\Gamma_a\rangle$ in Eq. \eqref{eq:BE} represents the thermally averaged total decay width of ALP. For $m_\Psi<m_a/2$, ALP can dominantly decay into two DM particles, resulting in an invisible decay width, given by
\beq
\Gamma (a\to \Psi\bar{\Psi})= c_\Psi^2\frac{m_a m_\Psi^2}{8\pi f_a^2}\sqrt{1-\frac{4m_\Psi^2}{m_a^2}}.
\label{eq:a-decay}
\eeq
Hence, to eliminate the double-counting from the on-shell ALP contribution in the $\Psi\bar{\Psi}\to\gamma\gamma$ process, a subtraction must be applied, which results in the corrected thermally averaged cross-section (used in Eq. \eqref{eq:BE}), expressed as \cite{Allen:2024ndv}: \beq
\langle\sigma_{\Psi\bar{\Psi}\to\gamma\gamma} v\rangle^{\rm sub}= \langle\sigma_{\Psi\bar{\Psi}\to\gamma\gamma} v\rangle - \frac{Y_a^{\rm eq}}{\mathcal{S}(Y_\Psi^{\rm eq})^2}\langle\Gamma_a\rangle~ {\rm BR}(a\to \Psi\bar{\Psi})~{\rm BR}(a\to\gamma\gamma).
\eeq
Although the decay channel (in Eq. \ref{eq:a-decay}) can significantly produce $\Psi$ at $T\sim m_a$, however, at $T\lesssim m_a$, the ALP abundance is Boltzmann suppressed and hence, this decay channel becomes inefficient. Moreover, since we consider the case of a thermal DM, which tracks its equilibrium abundance ($Y_\Psi = Y_\Psi^{\rm eq}$) before freeze out, the second term on the r.h.s of Eq. \ref{eq:BE} becomes negligible during that period.

On the other hand, the annihilation process $\Psi\bar{\Psi}\to\gamma\gamma$ produces a distinct gamma-ray line at $E_\gamma=m_\Psi$, which is subject to stringent constraints from gamma-ray line searches, including Fermi-LAT \cite{Albert:2014hwa,Foster:2022nva}, MAGIC \cite{MAGIC:2022acl}, HESS \cite{HESS:2018cbt}, EGRET, and COMPTEL \cite{Pullen:2006sy,Boddy:2015efa}. As pointed out in \cite{Dolan:2017osp,Acanfora:2023gzr, Allen:2024ndv}, achieving the correct DM relic density via this process while evading indirect detection limits at the same time is therefore highly challenging. To ensure $\Psi$ remains a viable freeze-out DM candidate with the observed relic abundance, $\Omega h^2\approx 0.12$ \cite{Planck:2018vyg}, $f_a$ typically needs to be at the $\mathcal{O}$(GeV) scale, which is further ruled out by collider searches \cite{Acanfora:2023gzr}.

Interestingly, a possible way to circumvent these constraints while achieving the correct relic abundance is to consider the resonance region, ${\rm where}~m_a/m_\Psi\approx 2$ \cite{Dolan:2017osp,Acanfora:2023gzr, Allen:2024ndv}. This allows for a smaller SM-ALP coupling (corresponds to a larger $f_a$). Consequently, Eq. \eqref{eq:svavg} exhibits a sharp peak around $s\approx m_a^2$ \cite{Dolan:2017osp}, indicating that the intermediate ALP is produced on-shell. It is convenient to use the narrow width approximation near the resonance region (with $\Gamma_a\ll m_a$), under which the ALP propagator can be effectively replaced by a Dirac delta function, simplifying the analysis as
\beq
\frac{1}{(s-m_a^2)^2+m_a^2\Gamma_a^2}\to \frac{\pi}{m_a\Gamma_a}\delta (s-m_a^2),
\eeq
which leads to an approximated expression for $\langle\sigma_{\Psi\bar{\Psi}\to\gamma\gamma} v\rangle$ as
\beq
\langle\sigma_{\Psi\bar{\Psi}\to\gamma\gamma} v\rangle\approx\frac{\pi}{128} \frac{ x g_{a\gamma\gamma}^2  r^5  K_1(x~r)}{  K_2(x)^2}.
\label{eq:svres}
\eeq
This result follows the assumption that $\Gamma (a\to \Psi\bar{\Psi})\gg\Gamma (a\to\gamma\gamma)$, such that the total ALP decay width can be approximated as $\Gamma_a\approx \Gamma (a\to \Psi\bar{\Psi})$, although the validity of this assumption depends on the choice of the mass ratio $r$ and $c_\Psi$. Requiring $\Gamma (a\to \Psi\bar{\Psi})\gg\Gamma (a\to\gamma\gamma)$ leads to the condition 
\beq
c_\Psi^2 \gg\frac{\alpha_{\rm EM}^2 }{8\pi^2}\frac{r^2}{\sqrt{1-4/r^2}}.
\label{cpsi}
\eeq 
As evident from Eq. \eqref{eq:svres}, near the resonance (while ensuring $m_a>2m_\Psi$), the annihilation rate depends primarily on $g_{a\gamma\gamma}$ and the mass ratio $r=m_a/m_\Psi$, while the DM-ALP coupling does not explicitly appear \footnote{However, this is not the case when $m_a<2m_\Psi~(r<2)$, as the decay $\Gamma (a\to \Psi\bar{\Psi})$ is forbidden when $r<2$.}.

\begin{figure}[htb!]
	$$
	\includegraphics[scale=0.46]{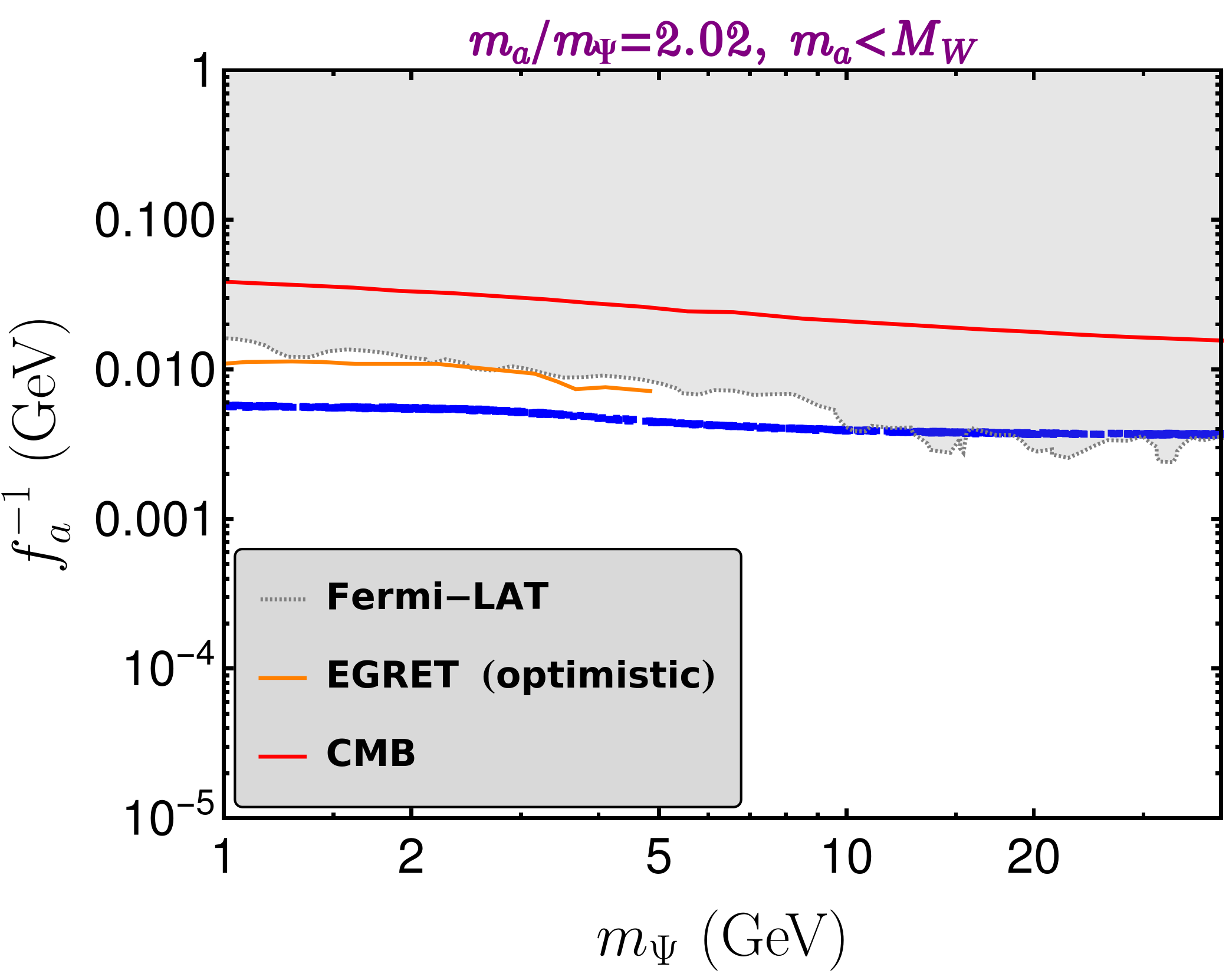}~
	\includegraphics[scale=0.4]{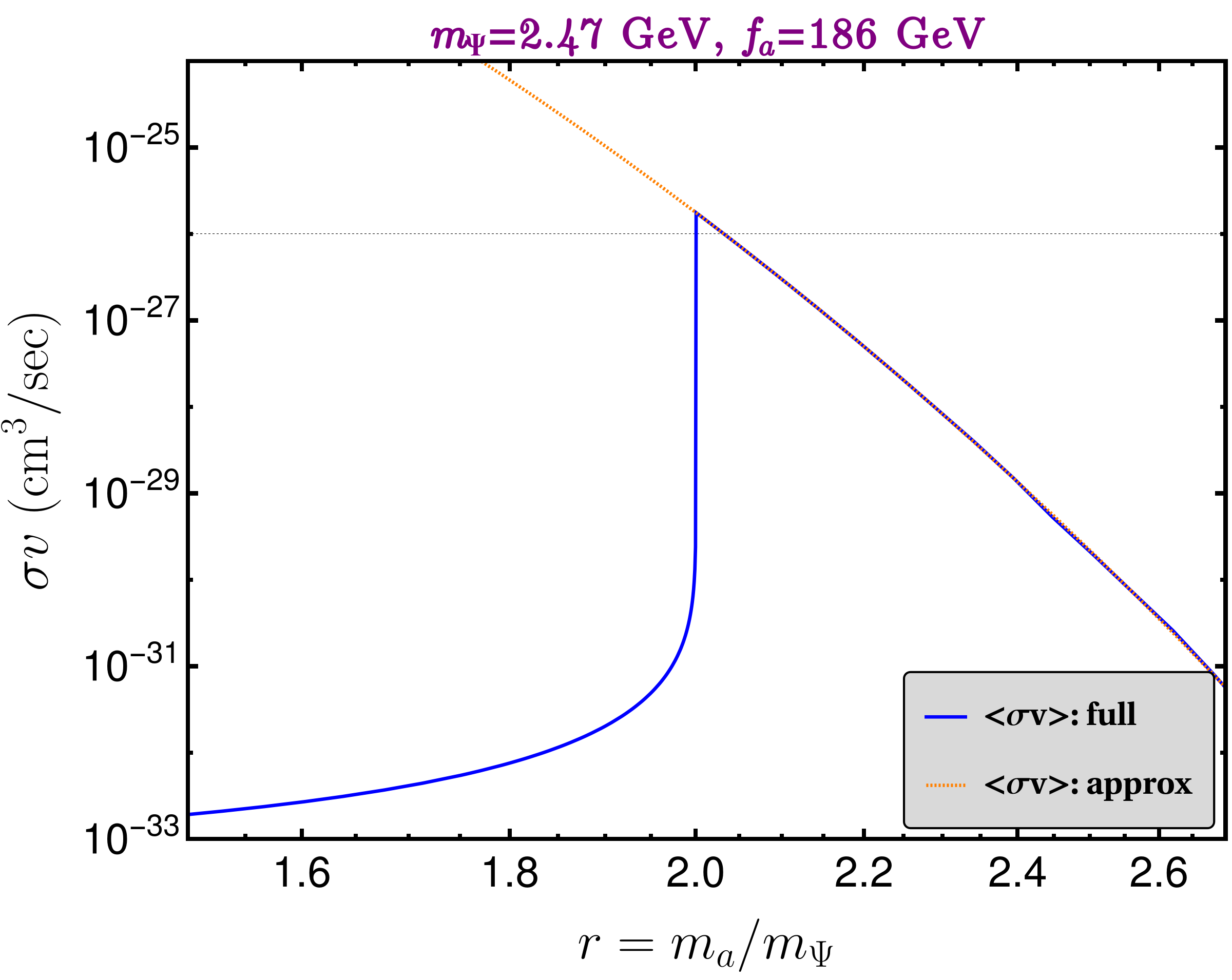}
	$$
	\caption{Left:  Relic satisfied DM parameter space (shown in blue contour line), with mass ratio $r=m_a/m_\Psi=2.02$. Also, constraints from CMB and indirect searches are shown, which are taken from \cite{Allen:2024ndv}. Right: Variation of thermally averaged cross section $\langle\sigma v\rangle$ against $r$ employing Eqs.~\eqref{eq:svavg}-\eqref{eq:sigma} (solid blue), while use of the approximated expression of Eq.~\eqref{eq:svres} is exhibited by dashed orange line (valid near resonance only). Here, $x=20$ is assumed, which is the typical DM mass to temperature ratio during DM freeze-out. The black grid line refers to $\langle\sigma v\rangle\sim \mathcal{O}(10^{-26}) ~{\rm cm}^3/{\rm s}$, needed for satisfying the correct relic abundance. }
	\label{fig:DM-ps}
\end{figure}

We then fix the mass ratio $r$ to a benchmark value $2.02$ and set $c_\Psi=1$, which is consistent with Eq.~\eqref{cpsi} since $r=2.02$ requires $c_\Psi\gg 4.7\times 10^{-3}$, and perform a parameter scan over $f_a$ (related to $g_{a\gamma\gamma}$ via Eq.~\eqref{eq:gagg}) and $m_\Psi$ to obtain a viable DM parameter space using {\tt MicrOmegas} \cite{Belanger:2018ccd}. Importantly,
for our benchmark choice $m_\Psi=2.47$ GeV, $r=2.02$, $f_a=186$ GeV and $c_\Psi=1$, we obtain $\Gamma(a\to \Psi \bar{\Psi})=4.9\times 10^{-6}~{\rm GeV}~~{\rm and}~~\Gamma(a\to \gamma\gamma)=1.09\times 10^{-10}~{\rm GeV}$, ensuring $\Gamma_a/m_a\ll 1$. This indicates the resonance is sufficiently narrow and the narrow width approximation is well justified. The values further indicates that the invisible decay channel dominates near the resonance despite the phase space suppression. Our analysis primarily focuses on the DM mass range $1~{\rm GeV}\lesssim m_\Psi<M_W/2$ GeV, motivated from the collider prospects. The result is shown in Fig.~\ref{fig:DM-ps} (left panel) where the blue line represents the relic density satisfied contour, while the grey-shaded region indicates the excluded portion by indirect detection. We find that for the choice of $r=2.02$, $m_\Psi\lesssim 10$ GeV, along with $f_a\sim \mathcal{O}(200)$ GeV ($g_{a\gamma\gamma}\sim \mathcal{O}(10^{-5})$ GeV$^{-1}$) can successfully reproduce the correct relic density via thermal freeze-out while remaining consistent with indirect detection bounds. It must be noted that other choices of mass ratios $(r)$ close to the resonance can also satisfy the correct relic density, as shown in Fig.~\ref{DM-colormap}.
\begin{figure}[htb!]
	$$
	\includegraphics[scale=0.6]{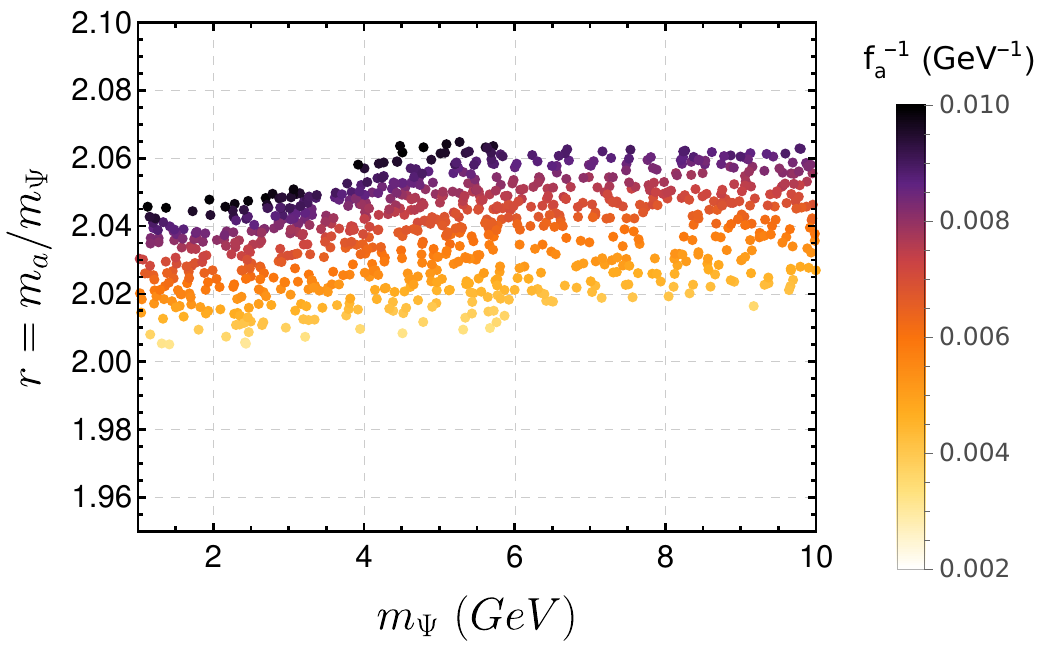}~
	$$
	\caption{Viable DM parameter space for mass ratios $r$ close to resonance in the DM mass range $1-10$ GeV. The color-bar shows the range of $f_a^{-1}$, required for relic satisfaction.}
	\label{DM-colormap}
\end{figure}
In the right panel of Fig.~\ref{fig:DM-ps}, we show the thermally averaged cross section for a benchmark point within the allowed parameter space $(m_\Psi = 2.47~ {\rm GeV}, ~f_a=186 ~{\rm GeV})$, against the mass ratio $r$. As shown by the solid blue and dashed orange curves, obtaining the correct relic abundance through thermal freeze-out requires staying close to the resonance region, {$i.e.$ with $r$ close to 2.02 justifying the choice of the benchmark value for $r$ in left panel figure}. In this region, the cross section reaches $\langle\sigma v\rangle\sim \mathcal{O}(10^{-26}) ~{\rm cm}^3/{\rm s}$, which is needed to reproduce the observed DM density. Notably, in the case of QCD axion mediation (instead of ALP), the required small values of $f_a$ would be ruled out by SN1987A, which places a stringent bound on KSVZ axion as $f_a\gtrsim 4\times 10^8$ GeV (corresponding to $m_a\lesssim \mathcal{O}(10^{-2})$ eV) \cite{Raffelt:2006cw,Caputo:2024oqc}. These supernova bounds apply only to ALP masses below a few hundred MeV. 

It is pertinent here to mention that the analysis so far relies on a crucial assumption that the DM remains in kinetic equilibrium during chemical freeze-out. Such assumption is generally justified in presence of large coupling between DM and the visible sector, ensuring efficient scattering between the DM and visible sector particles. However, if these scatterings become ineffective too early, DM can kinetically decouple before chemical freeze-out, altering the relic abundance. In particular, relatively weaker couplings in the resonance regime can give rise to early kinetic decoupling. In the next subsection, we examine how this effect alters the results compared to the benchmark values considered so far.

%%%%%%%%%%%%%%%%%%%%%%%%%%%%%%%%%%%%%%%%%%%%%%%%%%%%%%%%%%%%%%%%%%%%%%%%%%%%%%%%%%%%
\subsection{Effect of early kinetic decoupling in the resonance regime}
%%%%%%%%%%%%%%%%%%%%%%%%%%%%%%%%%%%%%%%%%%%%%%%%%%%%%%%%%%%%%%%%%%%%%%%%%%%%%%%%%%%%
In our case, the EKD effect is relevant in the ALP-resonance regime $(m_a\approx 2m_\Psi)$, where the annihilation cross section is enhanced by the resonance and consequently, the relic density can be achieved with larger values of $f_a$. Since the scatterings of DM with the thermal bath particles are not resonance-enhanced, they may fall out of equilibrium earlier, leading to early kinetic decoupling and a modified relic abundance.

\begin{figure}[h!]
	\centering
	\begin{tikzpicture}
		\begin{feynman}
			\vertex (a) at (0,1.5) {$\Psi$};
			\vertex (b) at (0,-1.5) {$a$};
			\vertex (c) at (2,0);
			\vertex (m) at (4.2,0);
			\vertex (d) at (6,1.5) {$\Psi$};
			\vertex (e) at (6,-1.5) {$a$};
			\diagram*{
				(a) -- [fermion, ultra thick] (c),
				(b) -- [scalar, ultra thick] (c),
				(c) -- [fermion, ultra thick, edge label'=$\Psi$] (m),
				(m) -- [fermion, ultra thick] (d),
				(m) -- [scalar, ultra thick] (e)
			};
			\vertex at (c) [blob, minimum size=0.5cm, fill=gray!50] {};
			\vertex at (m) [blob, minimum size=0.5cm, fill=gray!50] {};
		\end{feynman}
	\end{tikzpicture}
		\begin{tikzpicture}
		\begin{feynman}
			\vertex (a) at (0,1.5) {$\Psi$};
			\vertex (b) at (0,-1.5) {$a$};
			\vertex (c) at (3,1.5);
			\vertex (d) at (3,-1.5);
			\vertex (e) at (6,1.5) {$a$};
			\vertex (f) at (6,-1.5) {$\Psi$};
			\diagram*{
				(a) -- [fermion, ultra thick] (c) -- [scalar, ultra thick] (e),
				(b) -- [scalar, ultra thick] (d) -- [fermion, ultra thick] (f),
				(c) -- [fermion, ultra thick,edge label'=$\Psi$] (d)
			};
			\vertex at (c) [blob, minimum size=0.5cm, fill=gray!50] {};
			\vertex at (d) [blob, minimum size=0.5cm, fill=gray!50] {};
		\end{feynman}
	\end{tikzpicture}
	\begin{tikzpicture}
		\begin{feynman}
			\vertex (a) at (0, 1.5) {$\Psi$};
			\vertex (b) at (0, -1.5) {$V$};
			\vertex (c) at (3, 1.5);
			\vertex (d) at (3, -1.5);
			\vertex (e) at (6, 1.5) {$\Psi$};
			\vertex (f) at (6, -1.5) {$V$};
			\diagram* {
				(a) -- [fermion, ultra thick] (c) -- [fermion, ultra thick] (e),
				(b) -- [boson, thick] (d) -- [boson,  thick] (f),
				(c) -- [scalar, ultra thick, edge label'=$a$] (d)
			};
			\vertex at (c) [blob, minimum size=0.5cm, fill=gray!50] {};
			\vertex at (d) [blob, minimum size=0.5cm, fill=gray!50] {};
		\end{feynman}
	\end{tikzpicture}
	\caption{Upper panel: Feynman diagrams for $\Psi a \to \Psi a$ scattering processes (mediated by $\Psi$) in $s$ and $t$ channels. Lower panel: Feynman diagrams for $\Psi V \to \Psi V$ scattering processes (mediated by $a$) in $t$ channel, where $V$ stands for the SM gauge bosons: $\gamma, Z$ and $W$.}
	\label{scattering2}
\end{figure}
There are two main DM scattering channels present in our setup: {(I)} $\Psi a \to \Psi a$ (upper panel of Fig.~\ref{scattering2}) and {(II)} $\Psi V \to \Psi V$ (lower panel of Fig.~\ref{scattering2}), where $V\in \gamma, Z$ and $W$. Among these, the channel $\Psi a \to \Psi a$ plays the most significant role due to the presence of additional $s$-channel (left upper panel of Fig.~\ref{scattering2}). This enhances the interaction and allows the DM to remain in kinetic equilibrium for a longer duration. To illustrate the EKD effect on DM relic abundance, we adopt the previously considered benchmark parameter values: $m_\Psi=2.47~{\rm GeV}, ~f_a=186~{\rm GeV}$ and $r=2.02$ and compute the relevant reactions rates of DM. 
\begin{figure}[h!]
	$$
\includegraphics[scale=0.6]{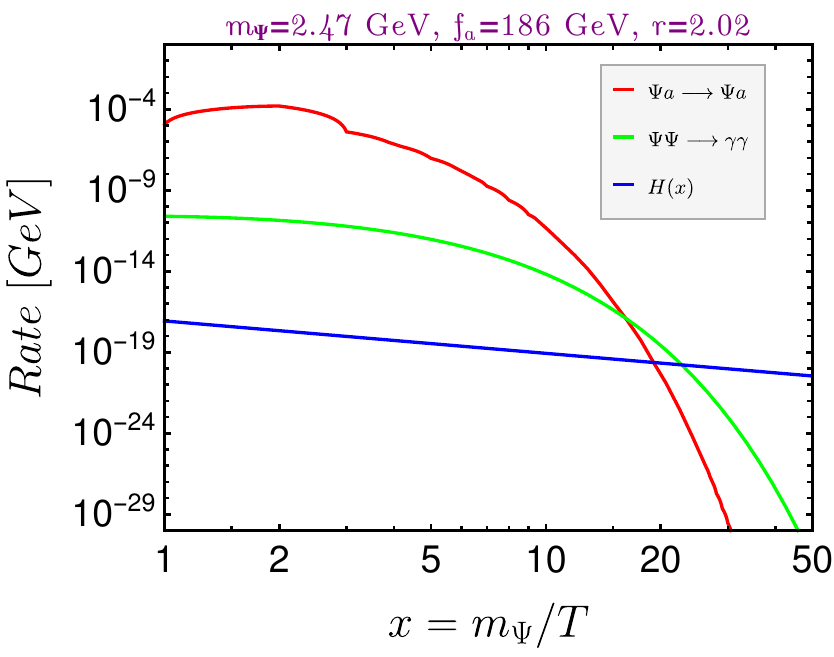}
$$
\caption{Different interaction rates of DM (solid red for $\Psi a \to \Psi a$ and solid green for $\Psi \Psi \to \gamma\gamma$) and the Hubble expansion rate of the Universe (solid blue) against $x=m_\Psi/T$. }
\label{reaction-rate}
\end{figure}
As shown in Fig.~\ref{reaction-rate}, the solid red curve represents the scattering (momentum transfer) rate for the process $\Psi a \to \Psi a$, which decouples slightly earlier than the DM annihilation rate $\Psi \Psi \to \gamma\gamma$ ($\sim n_\Psi^{\rm eq}\langle\sigma v\rangle$, indicated by the solid green curve). The value of $x$ where the condition $\Gamma_{\Psi a\to \Psi a}=H$ takes place is denoted by $x_{\rm kd}$. Beyond this point, the DM stops exchanging the momentum with the thermal bath and one can define a DM temperature $(T_\Psi)$ by relating it to the average kinetic energy, which evolves distinctly from the bath temperature $T$. Therefore, the DM temperature $T_\Psi$ can be defined as
\beq
T_\Psi = 
\begin{cases}
	T, & x \leq x_{\rm kd} \\
	\frac{m_\Psi}{x_{\rm kd}}\left(\frac{x}{x_{\rm kd}}\right)^2,  & x > x_{\rm kd}.
\end{cases}
\eeq
Such modification of the DM temperature alters the thermally averaged annihilation cross section (Eq.~\ref{eq:svavg}) at $x>x_{\rm kd}$ as
\beq
\langle\sigma_{\Psi\bar{\Psi}\to\gamma\gamma} v\rangle_{T_\Psi}=	\frac{1}{8m_\Psi^4 T_\Psi K_2^2(m_\Psi/T_\Psi)}
\int_{4m_\Psi^2}^{\infty}\sigma_{\Psi\bar{\Psi}\to\gamma\gamma} (s-4m_\Psi^2)\sqrt{s}K_1\left( { \sqrt{s}}/T_\Psi \right)  ds.
\label{eq:svavg2}
\eeq
Accordingly, post kinetic decoupling, the DM evolves via following modified Boltzmann equation\footnote{A more rigorous treatment of EKD requires solving the coupled Boltzmann equations for the DM number density and temperature, see e.g.~\cite{Bringmann:2006mu,Binder:2017rgn,Abe:2020obo,Duan:2024urq,Ding:2025eqq}.}:
\bea
\frac{dY_\Psi}{dx}&=& 
-\frac{\mathcal{S}(x)}{x H(x)}\left[\langle\sigma_{\Psi\bar{\Psi}\to\gamma\gamma} v\rangle_{T_\Psi} (Y_\Psi^2) - \langle\sigma_{\Psi\bar{\Psi}\to\gamma\gamma} v\rangle (Y_\Psi^{\rm eq})^2\right].
\label{eq:BE2}
\eea
In Fig.~\ref{fig:DM-evolution}, we present a comparison between the scenarios with and without EKD effects using the benchmark values of $m_\Psi$ and $f_a$. 
\begin{figure}[htb!]
	$$
	\includegraphics[scale=0.55]{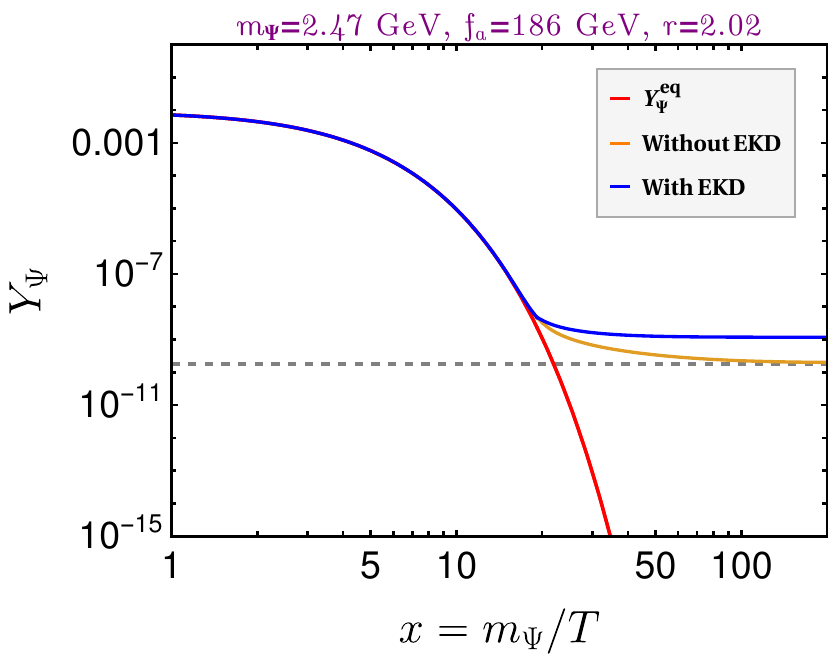}~
	\includegraphics[scale=0.55]{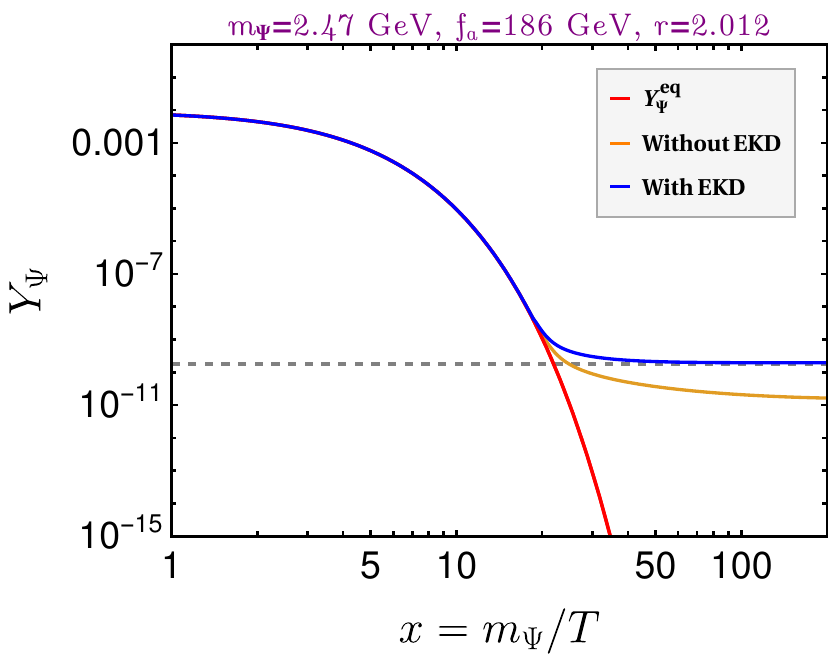}
	$$
	\caption{Evolution of DM with (sold blue) and without (solid orange) the early kinetic decoupling effect with $r=2.02$ (left panel) and $r=2.012$ (right panel). }
	\label{fig:DM-evolution}
\end{figure}
For $r=2.02$ (left panel), we observe a mild early kinetic decoupling (as demonstrated in Fig.~\ref{reaction-rate}) leads to an early DM freeze-out (solid blue line), resulting an overabundance of DM ($\sim 5$ times than the standard case). This overproduction can however be alleviated by varying $r$ or other parameters of the model. As shown in the right panel of Fig.~\ref{fig:DM-evolution}, choosing a slightly smaller ratio, $r=2.012$, allows the relic abundance to match the observed value even after including EKD effects. Thus, the EKD effect pushes the scenario closer to resonance (a reduction in $r$ of roughly $0.4\%$), though this shift does not significantly affect the collider analysis of the setup. In the next section, we turn to the collider implications of our setup.

%%%%%%%%%%%%%%%%%%%%%%%%%%%%%%%%%%%%%%%%%%%%%%%%%%%%%%%%%%%
\section{Event analysis at the electron-positron colliders}
\label{sec:col}
%%%%%%%%%%%%%%%%%%%%%%%%%%%%%%%%%%%%%%%%%%%%%%%%%%%%%%%%%%%%
High-energy electron-positron colliders provide a unique opportunity for comprehensive DM searches through missing energy associated with a visible particle. 
The channel that we focus here for ALP search is the mono-photon plus missing energy final state signature. In collider experiments, DM particles do not interact with the detector, rendering them invisible to direct observation. Consequently, they escape detection, appearing as `missing energy' in the recorded measurements. This `missing energy' serves as an indirect observation of DM production at the colliders \cite{ Yu:2013aca, Essig:2013vha, Kadota:2014mea, Yu:2014ula,Freitas:2014jla, Dutta:2017ljq, Choudhury:2019sxt, Horigome:2021qof, Barman:2021hhg, Kundu:2021cmo, Bhattacharya:2022qck,Ge:2023wye,Barman:2024nhr,Borah:2024twm,Barman:2024tjt,Borah:2025ema}\footnote{This should be contrasted with missing transverse momentum or colloquially missing transverse energy ($\slashed{E}_T$) accessible at hadron colliders, in absence of the information about the sub process CM energy of the colliding partons.}. The final state signal of our interest is associate ALP production with photon followed by the decay of ALP to DM pair as shown in Fig.~\ref{fig:alp.prod} \footnote{A $Z$-mediated $s$-channel contribution could also be possible if $C_{aB}\neq C_{aW}$, since $g_{a\gamma Z}$ becomes non-zero in that case.}. This type of associate mono-photon production in contrast to ISR-photon often referred as {\it natural mono-photon}, which proves to be beneficial for improving signal-background estimation, as we discuss below. The non-interfering irreducible SM background consists of two main contributions. One comes from a $W$ boson-mediated $t$-channel neutrino pair production, where a photon is radiated either from the initial state or from the $W$ boson (right plot of Fig.~\ref{fig:monoX.sm}). The other arises from $Z$-mediated $s$-channel neutrino pair production, with a photon emitted from the initial state (left plot of Fig.~\ref{fig:monoX.sm}). Other reducible SM backgrounds arise from the processes $e^+e^- \to X + \gamma$ where $X=\slashed{\gamma}, \slashed{\gamma} \slashed{\gamma}$ and $\slashed{e}^+\slashed{e}^-$. These contributions arise when the particles in $X$ is/are sufficiently soft or emitted in the forward direction such that they escape detection. As, the ILC detectors are assumed to be symmetric, therefore $\slashed{\gamma} \gamma$ background has negligible effect. In this analysis, we consider $\sqrt{s}=1$ TeV ILC with an integrated luminosity of $\mathfrak{L}_{\text{int}} = 5 \, \text{ab}^{-1}$ to perform a cut-based analysis. The DM phenomenology allowed benchmark point (BP) used in our study corresponds to a ALP mass $m_a = 5 \, \text{GeV}$ and a decay constant $f_a = 186 \, \text{GeV}$. This choice of $f_a$ results in an effective scale of order $\Lambda_{\text{eff}} \sim 4 \pi f_a \simeq 2.3 \, \text{TeV}$. This should be understood as a conservative upper bound rather than a strict equality, since the precise value of $\Lambda_{\text{eff}}$ is model dependent and is determined by the masses and couplings of the heavy degrees of freedom of a UV theory. For the benchmark values considered here, $\sqrt{s},m_a, m_\Psi<\Lambda_{\text{eff}}$ is satisfied, ensuring the validity of the EFT description in the context of collider.

\begin{figure}[htb!]
	\centering
	{\begin{tikzpicture}[baseline={(current bounding box.center),style={scale=0.7,transform shape}}]
			\begin{feynman}
			\vertex (a);
			\vertex [above left=2.4cm of a] (b) {\large $e^{+}$};
			\vertex [below left=2.4cm of a] (c) {\large $e^{-}$};
            \vertex[blob, fill=gray!60, minimum size=0.6cm] [right=1.8cm of a] (d) {};
			\vertex [above right=2.7cm of d] (e) {\large $\gamma$};
			\vertex[blob, fill=gray!60, minimum size=0.6cm] [below right=2.7cm of d] (f){};
			\vertex [above right=2cm of f] (g) {\large $\Psi$};
			\vertex [below right=2cm of f] (h) {\large $\bar{\Psi}$};
			\diagram{
					(b) -- [anti fermion, ultra thick, arrow size=2pt] (a) -- [ anti fermion, ultra thick, arrow size=2pt] (c);
					(a) -- [ ultra thick, boson, edge label={\large $\gamma$}] (d);
					(d) -- [boson, ultra thick] (e);
					(d) -- [scalar, ultra thick, edge label={\large $a$}] (f);
					(f) -- [fermion, ultra thick, arrow size=2pt] (g);
					(f) -- [anti fermion, ultra thick, arrow size=2pt] (h);
			};
			\end{feynman}
	\end{tikzpicture}}	
	\caption{Feynman diagrams that induce mono-$\gamma$ + missing energy final state within EFT framework from ALP 
	production. }
	\label{fig:alp.prod}
\end{figure}
\begin{figure}
		\begin{tikzpicture}
		\begin{feynman}
			\vertex (a) at (0, 1.7) {\Large $e^-$};
			\vertex (b) at (0, -1.7) {\Large $e^+$};
			\vertex (c) at (2.3, 0);
			\vertex (c1) at (4.7, 0);
			\vertex (d) at (7, 1.7) {\Large $\nu$};
			\vertex (e) at (7, -1.7) {\Large $\bar{\nu}$};
			
			\diagram*{
				(a) -- [fermion, ultra thick, arrow size=2pt] (c) -- [fermion, ultra thick, arrow size=2pt] (b),
				(e) -- [fermion, ultra thick, arrow size=2pt] (c1) -- [fermion, ultra thick, arrow size=2pt] (d),
				(c) -- [boson, ultra thick, edge label=$Z$] (c1)
			};
		\end{feynman}
	\end{tikzpicture}~~~
		\begin{tikzpicture}
		\begin{feynman}
			\vertex (a) at (-1, 1.6) {\Large $e^-$};
			\vertex (b) at (-1, -1.6) {\Large $e^+$};
			\vertex (c) at (2, 1.6);
			\vertex (d) at (2, -1.6);
			\vertex (e) at (5, 1.6) {\Large $\nu$};
			\vertex (f) at (5, -1.6) {\Large $\bar{\nu}$};
			
			\diagram* {
				(a) -- [fermion, ultra thick, arrow size=2pt] (c) -- [fermion, ultra thick, arrow size=2pt] (e),
				(b) -- [anti fermion, ultra thick, arrow size=2pt] (d) -- [anti fermion, ultra thick, arrow size=2pt] (f),
				(c) -- [boson, ultra thick, edge label=$W$] (d)
			};
		\end{feynman}
	\end{tikzpicture}
	\caption{Feynman diagrams of non interfering SM backgrounds contributing to mono-$\gamma$+ $\slashed{E}$ final state at the $e^+e^-$ colliders.  Left: $Z$ mediated $s$-channel diagram where a photon can be radiated from initial states, contributing to the resulting final state;  right: $W$ mediated $t$-channel diagram where a photon can arise both from the initial states and from the $W$-boson.}
	\label{fig:monoX.sm}
\end{figure}

We generate signal and background events in $\tt Madgraph$ \cite{Alwall:2014hca}, then showered and analyzed through {\tt Pythia} \cite{Sjostrand:2014zea} and the detector simulation is done by ${\tt Delphes}$ \cite{deFavereau:2013fsa} using {\tt ILCgen} card though parametrized resolution functions. These smearing prescriptions are designed to emulate the expected performance of the ILC detectors and provide a fast and efficient approximation to a full detector simulation. The $\tt UFO$ file that is feeded to {\tt Madgraph} is generated through $\tt FeynRules$ \cite{Christensen:2008py}. During event generation, we constrain the phase space by requiring the photon to have a transverse momentum of $p_T^\gamma > 10$ GeV and a pseudorapidity within $|\eta_{\gamma}| \leq 2.5$. With these selection cuts, soft and collinear photons are effectively eliminated, making the reducible backgrounds negligible. Additionally, we consider only events featuring a single photon explicitly in the final state while ensuring no leptons or jets are present. The key variables for cut-based analysis relevant to the mono-$\gamma$ signal are  defined as follows:
\begin{itemize}
	\item \textbf{Missing energy ($E_{\text{miss}}$):} The energy carried away by missing particle is known as missing energy which is defined from the knowledge of $\sqrt{s}$ as
\beq
E_{\text{miss}}=\sqrt{s}-\sum_{i}E_{i},
\eeq
where $i$ runs over all the visible particles in the final state.

\item \textbf{Pseudorapidity ($\eta_{\gamma}$):} The definition of $\eta_{\gamma}$ is given by 
\beq
\eta_{\gamma}=\frac{1}{2}\log\left[\frac{E_{\gamma}+p_z}{E_{\gamma}-p_z}\right],
\eeq
where $E_{\gamma}$ and $p_z$ are the energy and $z$ component of 4-momenta of the ISR photon.
\end{itemize}
\begin{figure}[htb!]
	%	\centering
	$$
	\includegraphics[scale=0.4]{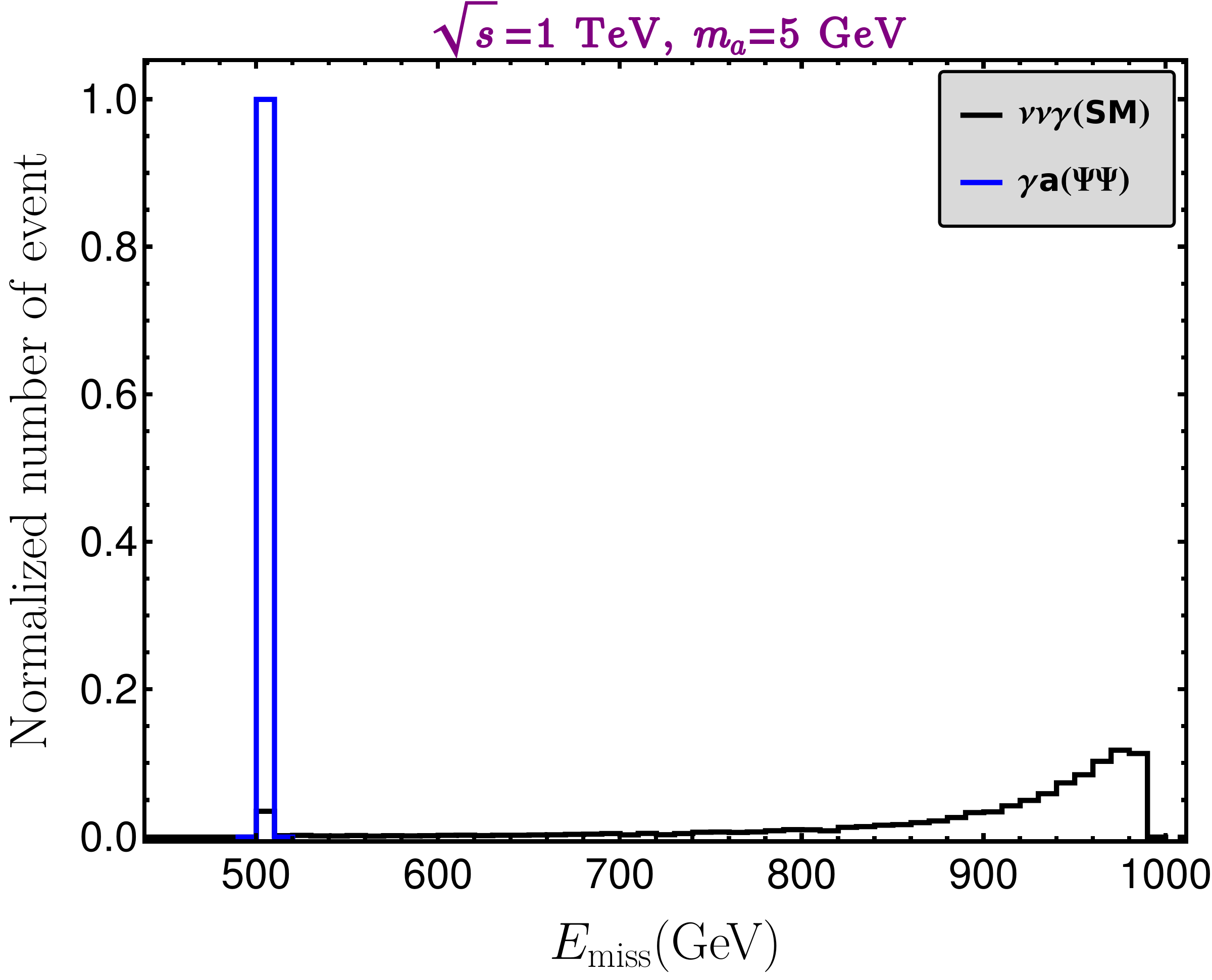}~~
	\includegraphics[scale=0.4]{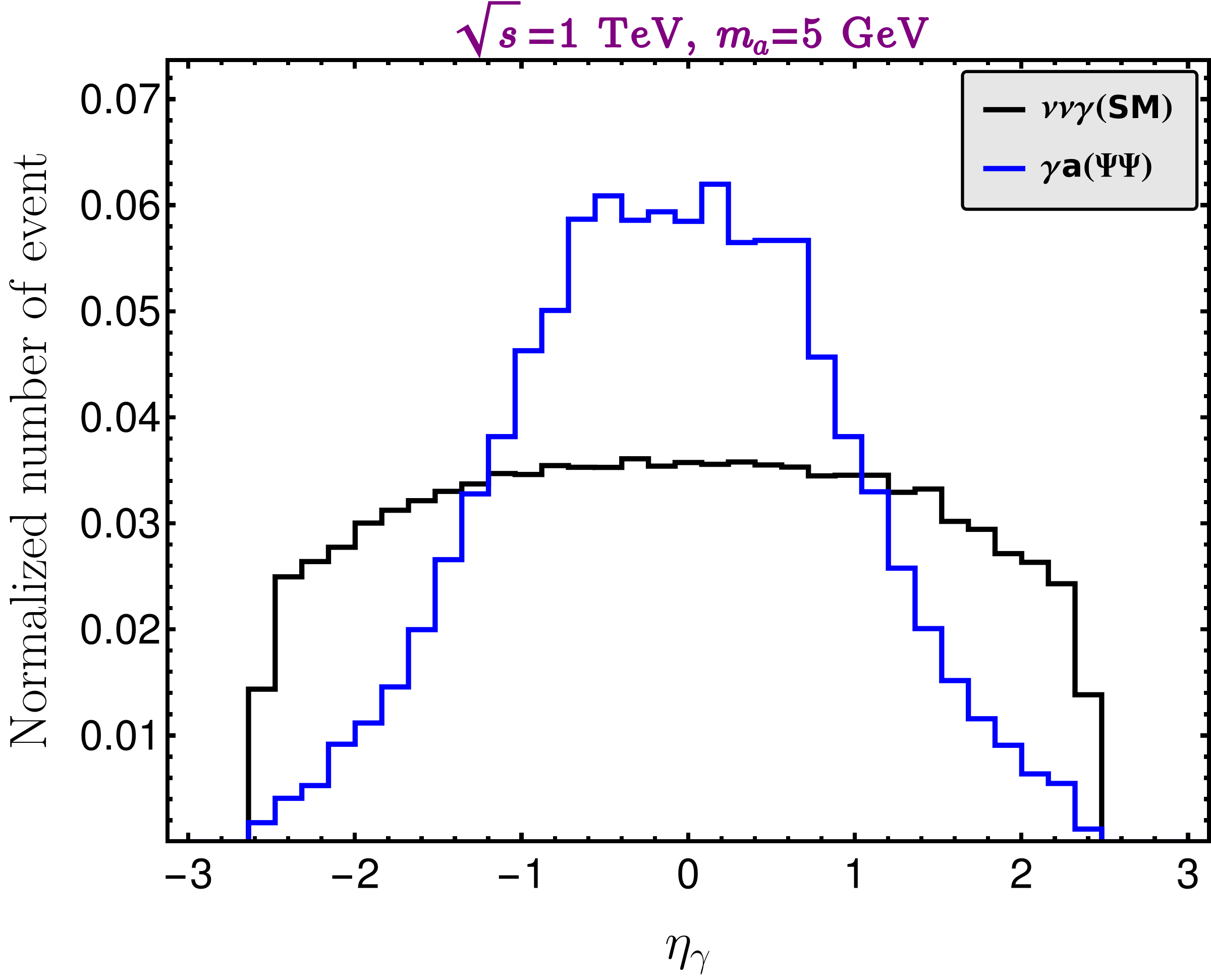}
	$$
	\caption{Normalized event distribution of kinematic variables at the ILC with $\sqrt{s}=1$ TeV and $m_a=5$ GeV. Left: missing energy ($E_{\text{miss}}$), right: pseudorapidity ($\eta_{\gamma}$). }
	\label{fig:event.dist}
\end{figure}

The missing energy distribution of the SM background exhibits a distinctive double-peak structure, as shown in the left plot of Fig.~\ref{fig:event.dist}. The first peak, occurring around 1000 GeV, is primarily arise to the dominant $W$-mediated $t$-channel process. Meanwhile, a secondary, sub-dominant peak appears near 500 GeV, resulting from the $Z$-mediated $s$-channel process. The position of the secondary peak at the tail of the distribution can be determined using the following expression
\beq
E_{\text{miss}}=\frac{\sqrt{s}}{2}\left(1+\frac{m_Z^2}{s}\right),
\eeq
where $m_Z$ is the $Z$ boson mass. On the other hand, it is quite understandable that the signal distribution peaks at 500 GeV,  as $E_{\text{miss}}=\sqrt{s}-E_\gamma$ here, and given that ALP mass is chosen to be very small (almost massless compared to the CM energy), the available energy is roughly divided into half and half between the axion and photon\footnote{In the case of the off-shell decay of the ALP into a DM pair, the event distribution of $E_{\text{miss}}$ exhibits a falling nature, peaking around 500 GeV and gradually decreasing at higher energies.}.  We must reiterate here that the spectacular distinction between the signal and background missing energy distribution in mono-photon channel results from producing the photon in the final state (see Fig.~\ref{fig:alp.prod}), and not as an initial state radiation (ISR), as is the case with most of the DM scenarios. This feature in turn, stems from the ALP coupling with the field strength tensor as in Eq.~\eqref{eq:ALP-SM2}. Therefore, upper cut on $E_{\text{miss}}$ eliminates 97\% background keeping the signal intact. Furthermore, applying an absolute pseudorapidity (distribution shown in right plot of Fig.~\ref{fig:event.dist}) cut of $|\eta_{\gamma}| > 1$ reduces the background by approximately 33\%, while the signal is reduced by only about 12\%. Linear electron-positron colliders possess partially polarized electron and positron beams which in advantageous in reducing SM backgrounds. Within the SM, left-handed leptons have stronger couplings strength than the right-handed leptons with $Z$-boson. Thus, right polarized electron and left-polarized positron beam reduce the SM background processes that contribute to the mono-photon final state. Based on the ILC snowmass report \cite{CMB-HD:2022bsz}, we choose $\{P_{e^+}:P_{e^-}\}=\{-20\%:+80\%\}$ combination that offers a five-fold suppression of the SM while amplifying the signal by 16\% as apparent from Table~\ref{tab:cut.tab}.
%%%%%%%%%%%%%%%%%%%%%%
\begin{table}[h!]
	\centering
	\begin{tabular}{|c|c|c|c|c|c|c|}
		\hline
		\multirow{2}{*}{Cuts} & \multicolumn{3}{c|}{$\{P_{e^{+}}, P_{e^{-}}\} = \{0 \%, 0 \%\}$} & \multicolumn{3}{c|}{$\{P_{e^{+}}, P_{e^{-}}\} = \{-20 \%, +80 \%\}$} \\ \cline{2-7}
		& $\Psi \bar{\Psi} \gamma$ & $\nu \overline{\nu} \gamma$ & Significance & $\Psi \bar{\Psi} \gamma$ & $\nu \overline{\nu} \gamma$ & Significance \\ \hline
		Basic cuts & 1750 & 12440000 & 0.50 & 2050 & 2389000 & 1.32 \\
		$E_{\text{miss}} < 510$ GeV & 1750 & 435400 & 2.63 & 2050 & 83615 & 6.95 \\
		$|\eta_{\gamma}| > 1$ GeV & 1545 & 292589 & 2.86 & 1810 & 56189 & 7.64 \\
		\hline
	\end{tabular}
	\caption{Cutflow for signal and SM background events for mono-$\gamma$ signal at the ILC with $\sqrt{s}$ = 1 TeV and $\mathfrak{L}_{\text{int}}$ = 5 $\rm{ab^{-1}}$ for unpolarized ($\{P_{e^{+}}: P_{e^{-}}\} = \{0 \%: 0 \%\}$) and polarized ($\{P_{e^{+}}: P_{e^{-}}\} = \{-20 \%: +80 \%\}$) cases.}
	\label{tab:cut.tab}
\end{table}

\begin{figure}[H]
	%	\centering
	$$
	\includegraphics[scale=0.4]{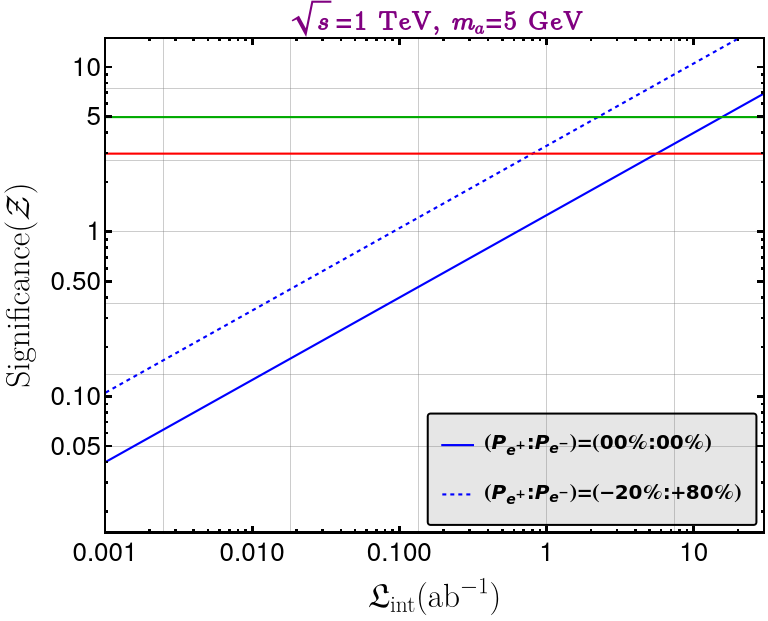}~~
	\includegraphics[scale=0.4]{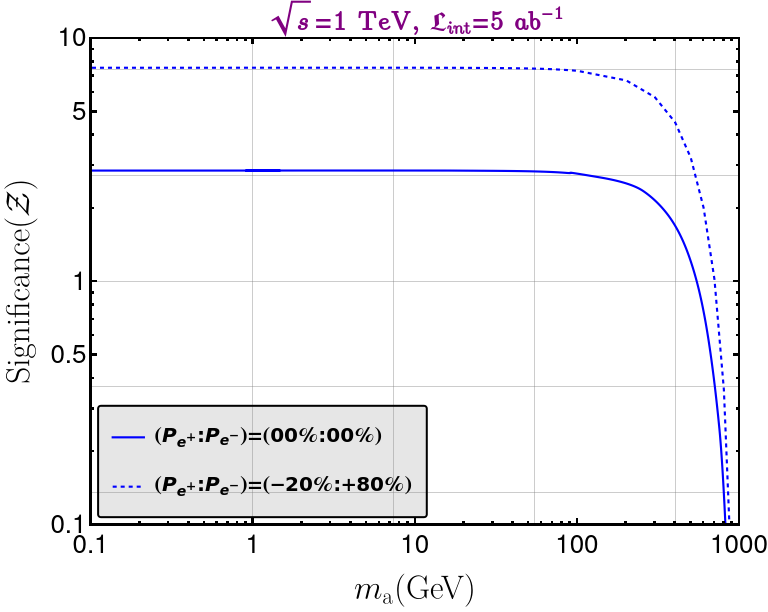}
	$$
	\caption{Left: variation of significance ($\mathcal{Z}$) with luminosity ($\mathfrak{L}_{\text{int}}$) for a fixed $m_a=5$ GeV for both unpolarized and polarized beams. Red (Green) line indicates 3$\sigma$ (5$\sigma$) significance; right: variation of $\mathcal{Z}$ with the ALP mass ($m_a$) for a fixed $\mathfrak{L}_{\text{int}}=5$ ab$^{-1}$ for both unpolarized and polarized beams. Notably, the ALP mass range considered here is allowed only from the CM energy reach of the ILC and not directly associated to DM phenomenology.}
	\label{fig:lum}
\end{figure} 
The signal significance ($\mathcal{Z}$) is defined as \cite{Cowan:2010js}
\beq
\mathcal{Z}=\sqrt{2\left[(S+B)\log\left(1+\frac{S}{B}\right)-S\right]},
\eeq
where $S$ and $B$ are the signal and background events, respectively. For the condition $B\gg S$, this expression reduced to $\mathcal{Z}=S/\sqrt{B}$. After the cuts mentioned in Table~\ref{tab:cut.tab}, $\mathcal{Z}$ is approximately 2.85$\sigma$ for unpolarized beams with $\sqrt{s}=1$ TeV and $\mathfrak{L}_{\text{int}}=5~\rm{ab^{-1}}$. With the beam polarization, the $\mathcal{Z}$ is enhanced to 7.64$\sigma$. In the left of Fig.~\ref{fig:lum}, we show the variation of $\mathcal{Z}$ with $\mathfrak{L}_{\text{int}}$ for both polarized and unpolarized beams. We find that a significance of 3$\sigma$ (5$\sigma$) can be achieved with an integrated luminosity of $\mathfrak{L}_{\text{int}} = 5.5\ (15.5)\ \text{ab}^{-1}$ for unpolarized beams, whereas, for polarized beams, the same level of significance can be reached with a significantly lower luminosity of $\mathfrak{L}_{\text{int}} = 0.8\ (2.2)\ \text{ab}^{-1}$. In the right panel of Fig.~\ref{fig:lum}, we present the variation of  $\mathcal{Z}$ as a function of $m_a$ for a fixed $\mathfrak{L}_{\text{int}}$. We notice that $\mathcal{Z}$ remains approximately constant up to $m_a = 100$ GeV, as the cross-section does not change significantly in this range. However, beyond $m_a = 100$ GeV, $\mathcal{Z}$ starts to decrease due to phase-space suppression. We note that a similar analysis can be performed at $\sqrt{s}=250~\mathrm{GeV}$ with $\mathfrak{L}_{\text{int}} = 2~\mathrm{ab}^{-1}$ (maximum reach of ILC250), and at $\sqrt{s}=500~\mathrm{GeV}$ with $\mathfrak{L}_{\text{int}} = 4~\mathrm{ab}^{-1}$ (maximum reach of ILC500). For ILC250, using the beam polarization considered here, the resulting $\mathcal{Z}$ remains below the $3\sigma$ exclusion limit, whereas ILC500 is capable of probing the benchmark point at the $5\sigma$ discovery level. If we push further to ILC1000, higher CM energy leads better background reduction resulting in an overall improvement in the $\mathcal{Z}$ which is the scenario considered here.  

It is important to reiterate that the signal observability as achieved here, relies on a few key factors, first the ALP-photon interaction, 
which makes the photon appearing out of the production vertex, and thus distinguishable from the initial state radiation photon as in the SM background via 
$E_{\rm miss}$ and $\eta_\gamma$ variables; second is the partial polarizability of the initial beams which reduces the SM background significantly, and enhancing the signal. 
Both of these can be achieved at the future electron-positron machine, while it is difficult to achieve the same at the present LHC or in its high luminosity projection as discussed in Appendix~\ref{sec:hl-lhc}. The opposite-sign muon (OSM) + $E_{\text{miss}}$ channel provides a complementary probe, especially effective at high CM energies. We therefore explore this channel at a 10 TeV muon collider \cite{InternationalMuonCollider:2025sys}, as discussed in Appendix \ref{sec:osms}.

%%%%%%%%%%%%%%%%%%%%%%%%%%%%%%%%%%%%
\section{Estimation of $g_{a \gamma \gamma}$}
\label{sec:coll.sen}
%%%%%%%%%%%%%%%%%%%%%%%%%%%%%%%%%%%%
The key parameter that allows us for collider detection of ALP is the ALP-photon coupling $g_{a\gamma\gamma}$. In this section, we present the estimation in accuracy measurement of $g_{a\gamma\gamma}$ coupling following the cut-based analysis discussed above. 
To achieve this, we adopt the conventional binned $\chi^2$ method which is defined as follows:
\begin{equation}
	\chi^2 = \sum_{j}^{\rm{bins}} \left( \frac{N_j^{\tt obs} - N_j^{\tt theo}(g_i)}{\Delta N_j} \right)^2,
	\label{eq:chi2.col}
\end{equation}
where $N_j^{\tt obs}$ denotes the number of simulated events in the $j^{\tt th}$ bin, and $N_j^{\tt theo}(g_i)$ represents the corresponding number of events predicted theoretically in the presence of NP, which depends on the coupling parameter(s) $g_i$. The summation extends over all $\cos \theta$ bins of the chosen observable, the differential scattering cross-section of the process $e^+e^- \to \gamma a(\chi \chi)$ defined in Eq.~\eqref{eq:diff.cross.ALP} in Appendix \ref{sec:diff.cross}, which is employed in the analysis. The statistical uncertainty in each bin, $\Delta N_j$, is taken to be $\sqrt{N_j^{\tt obs}}$, under the assumption that the event counts follow a Poisson distribution. 
\begin{figure}[htb!]
	%	\centering
	$$
	\includegraphics[scale=0.7]{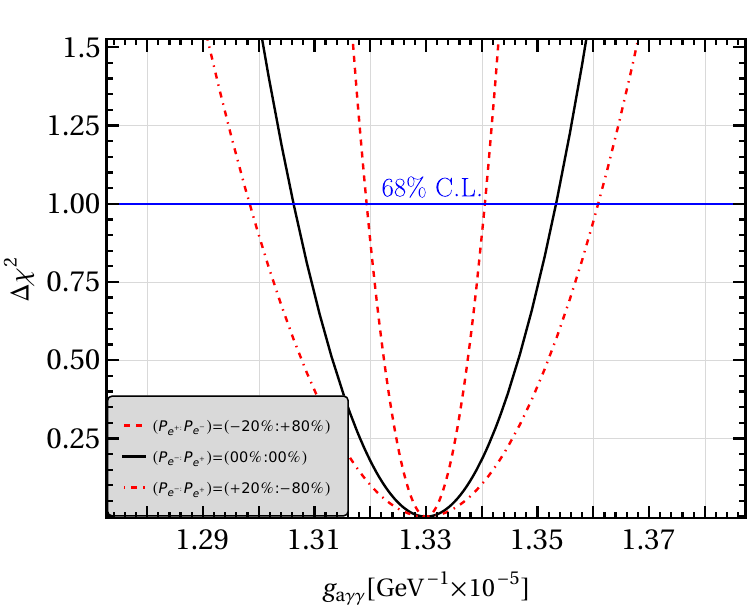}~
	$$
	\caption{$\Delta \chi^2$ variation with $g_{a\gamma \gamma}$ coupling at the ILC with $\sqrt{s}=1$ TeV, $\mathfrak{L}_{\rm{int}}=500$ fb$^{-1}$, and three different beam polarization written in the insect with $m_a=5$ GeV.}
	\label{fig:chi2}
\end{figure}
Our objective is to estimate the precision with which $g_{a\gamma \gamma}$ coupling can be estimated at the $e^+e^-$ colliders. We set the input of $g_{a\gamma \gamma}=1.33 \times 10^{-5}$ GeV$^{-1}$ with $m_a=5$ GeV in order to be consistent with  DM phenomenology. The variation of the $\Delta\chi^2=\chi^2-\chi^2_{\text{min}}$ function with respect to the NP coupling is illustrated in Fig.~\ref{fig:chi2}. Here, $\chi^2_{\text{min}}=0$}.  For unpolarized beam, the accuracy of $g_{a\gamma \gamma}$ coupling is approximately 1.87\%. If we consider $\{P_{e^+}:P_{e^-}\}=\{-20\%:+80\%\}$ polarization combination, non-interfering SM background is reduced substantially while the signal increases, the accuracy is enhanced to the 0.87\%. The `optimal' accuracy is obtained when the numerical bin-by-bin event distribution reaches close to that of the analytical distribution. This requires higher number of events and more luminosity. However, in the present context the betterment would be marginal, given the huge SM background left even after employing selection cuts.

\begin{figure}[htb!]
	%	\centering
	$$
	\includegraphics[scale=0.7]{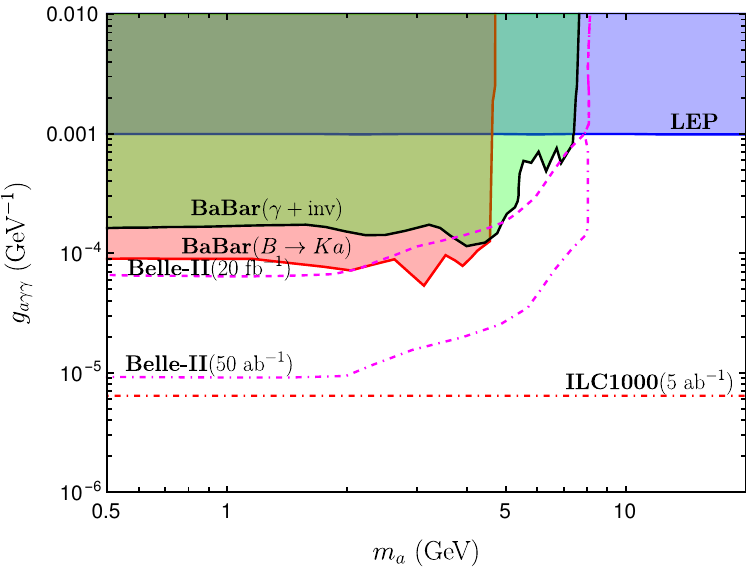}
	$$
	\caption{Summary of existing constraints at 95\% C.L. on $g_{a \gamma \gamma}-m_a$ plane from the mono-$\gamma$ + $E_{\text{miss}}$ search via electron-positron collison at the LEP \cite{DELPHI:2008uka,Fox:2011fx}, and BaBar \cite{Izaguirre:2016dfi}, $B \to K a$ with invisible decay of ALP at BaBar \cite{BaBar:2017tiz,Darme:2020sjf}. We also show the projected sensitivity at the Belle-II experiment with $\mathfrak{L}_{\text{int}}$ = 20 fb$^{-1}$ (magenta dashed) and 50 ab$^{-1}$ (magenta dot-dashed) \cite{Dolan:2017osp,Belle-II:2018jsg}.  Our result from ILC1000 with $\mathfrak{L}_{\text{int}}$ = 5 ab$^{-1}$ with beam polarization ($\{P_{e^+}:P_{e^-}\}=\{-20\%:+80\%\}$) is shown in red dot-dashed line.}
	\label{fig:constraints}
\end{figure}
At 95\% C.L., assuming $C_{aB}=C_{aW}$, the existing constraints and projected sensitivities on $g_{a\gamma\gamma}$ as a function of $m_a$ are shown in Fig.~\ref{fig:constraints}, derived from invisible ALP decay searches across the experimental frontiers mentioned above. For $m_a$ up to 20~GeV, the constraint on $g_{a\gamma \gamma}$ from the LEP via mono-$\gamma$ + $E_{\rm miss}$ search is approximately $\mathcal{O}(10^{-3})$ GeV$^{-1}$. Using the same final state in electron-positron collisions, the BaBar experiment sets a constraint of $\mathcal{O}(10^{-4})$ GeV$^{-1}$ for $m_a \lesssim 5$~GeV. The BaBar experiment has also provided an improved constraint, roughly a factor of 2 stronger than the former, through the process $B \to K a$ with invisible ALP decay. At the Belle-II, the sensitivity on $g_{a\gamma \gamma}$ is approximately $6 \times 10^{-5}$ ($1 \times 10^{-5}$) GeV$^{-1}$ for an integrated luminosity of $\mathfrak{L}_{\rm int} = 20$~fb$^{-1}$ (50~ab$^{-1}$) up to $m_a \sim 2$~GeV, after which it decreases due to phase-space suppression. Our results indicate that the sensitivity at the ILC1000 with polarized beams reaches roughly $6 \times 10^{-6}$ GeV$^{-1}$ at $\mathfrak{L}_{\rm int} = 5$~ab$^{-1}$. 
%%%%%%%%%%%%%%%%%%%%%%%%%%%%%%%%%%%
\section{Conclusions}
\label{sec:con}
%%%%%%%%%%%%%%%%%%%%%%%%%%%%%%%%%%%

In this work, we have examined the role of ALPs as a portal between a Dirac fermion DM and the SM sector, and investigated the prospects of probing such ALP-mediated DM at electron-positron colliders. This framework involves a derivative interaction between the DM and ALP, along with low-energy effective interactions of the ALP with SM electroweak gauge bosons. Focusing on the mass hierarchy $m_a\gtrsim 2 m_{\Psi}$ and $m_a<M_W, M_Z$, we demonstrated that for relatively lower values of the ALPs decay constant $f_a$, thermal DM production can be viable, with the dominant annihilation channel being $\Psi \Psi \to \gamma \gamma$, mediated by the ALP. We have showed that this annihilation channel can yield the correct DM relic abundance, while evading the strong indirect detection constraints, only in the near-resonance region where $m_a\approx 2 m_\Psi$. Taking a benchmark ratio between the DM mass and ALP mass, $r=m_a/m_\Psi=2.02$, we performed a detailed DM phenomenological analysis and found that DM masses $m_\Psi \lesssim \mathcal{O}(10~{\rm GeV})$ can successfully reproduce the observed relic density.  The presence of small coupling in the resonance region might be problematic for keeping the DM to be present in kinetic equilibrium via elastic scattering leading to early kinetic decoupling. The effect of such early kinetic decoupling has been investigated as well. In this setup, the invisible decay channel $a\to \Psi \Psi$ becomes kinematically allowed and actually dominates over the diphoton decay mode, making it a dominant channel at the colliders giving rise to missing energy signature.

Following the DM phenomenology, we have conducted a signal-background analysis considering the invisible decay of the ALP at the International Linear Collider (ILC). The study is performed at a CM energy of 1~TeV with an integrated luminosity of 5~ab$^{-1}$, considering initial beam polarization. The final state under investigation is characterized by a mono-photon plus missing energy. In this context, missing energy and pseudorapidity emerge as crucial kinematic variables, effectively suppressing substantial non-interfering SM  neutrino background. {This is possible due to the very specific ALP-photon interaction, which makes the mono photon signal largely segregable from that of the 
SM contamination. We must note here that most of the SM-DM effective interactions fail to provide this discrimination and hence should be noted carefully.}   
For the beam polarization combination $\{P_{e^+} : P_{e^-}\} = \{-20\% : +80\%\}$ at 1~TeV CM energy, we have obtained exclusion ($3\sigma$) and discovery ($5\sigma$) limits for integrated luminosities of 1~ab$^{-1}$ and 3~ab$^{-1}$, respectively. Furthermore, we have employed the standard $\chi^2$ analysis to estimate the sensitivity to the ALP-photon couplings under the specified collider configuration. With the aforementioned beam polarization, the accuracy in determining the coupling can reach approximately 1\%.

In summary, our work highlights a unique feature of ALP-mediated fermionic DM that enables its detection at future lepton colliders through a mono-photon plus missing-energy signature. When the ALP couples to field-strength tensors, the photon is emitted from the final state, producing a missing-energy spectrum that is clearly distinguishable from ISR photons (present in most of the DM scenarios) by angular-momentum considerations. This sharp feature, unlike typical mono-photon dark-matter searches dominated by ISR, allows the ALP signal to stand out even for very small production rates, making it possible to probe ALP-photon couplings as small as $\mathcal{O}(10^{-5}) ~{\rm{GeV}}^{-1}$ while remaining consistent with relic density. Such discrimination is not achievable at the HL-LHC, underscoring a unique advantage of lepton colliders and motivating this new direction for ALP searches.

\vspace{1.5cm}
\noindent
\hspace{0.1cm}{\Large \textbf{Acknowledgements}}\\
\vspace{0.001cm}\\
SB acknowledges the ANRF fund CRG/2023/000580 from the Govt. of India. SJ and SKM acknowledge financial support from CRG/2021/005080 project under SERB,
Govt. of India. The work of AS is supported by the grants CRG/2021/005080 and
MTR/2021/000774 from SERB, Govt. of India. SJ thanks Abhik Sarkar for useful discussions.

\appendix
%%%%%%%%%%%%%%%%%%%%%%%%%%%%%%%%%%%%%%%%%%%%%%%%%%%%%%%%%%%%%%%%%%%%%%%%%%%%%%%%
\section{UV Completion of ALP-DM Coupling Without ALP-SM Fermion Interactions}
\label{sec:uv.comp}
%%%%%%%%%%%%%%%%%%%%%%%%%%%%%%%%%%%%%%%%%%%%%%%%%%%%%%%%%%%%%%%%%%%%%%%%%%%%%%%%%
Here, we outline a minimal {UV-complete model} where ALP couples to a Dirac fermionic DM, while remaining decoupled from the SM fermions. To do so, we consider a global axial $U(1)$ symmetry that is spontaneously broken at a scale $f_a$ by the vacuum expectation value of a complex scalar field $\Phi$, which can be parameterized as
\bea
\Phi=\frac{1}{\sqrt{2}}(f_a+\rho) e^{ia/f_a}\,,
\eea
where $\rho$ is the massive radial degree of freedom and $a$ is the ALP, which emerges as the pNGB of the broken symmetry. We also consider a Dirac fermionic DM candidate $\Psi$ and heavy vector like fermions (VLF) $Q$, which are charged under $U(1)$ symmetry and transform as
\bea
\Phi \to e^{i\alpha}\Phi,~~~~\Psi \to e^{i(\alpha/2)\gamma_5}\Psi
,~~~~Q \to e^{i(\alpha/2)\gamma_5}Q \,.
\eea
Here, we further consider the VLFs $(Q)$ to be charged under SM electroweak symmetry as well. Importantly, the SM fermions are assumed to be singlets under $U(1)$, and thus do not couple directly to the ALP. Therefore, the relevant UV Lagrangian for the $U(1)$ sector is given by
\bea
\mathcal{L}\supset |\partial_\mu \Phi|^2-V(\Phi)+\overline{\Psi} i \slashed{\partial} \Psi+\bar{Q} i \slashed{\partial} Q  +(y_\Psi \Phi\overline{\Psi}_L\Psi_R + y_Q \Phi\overline{Q}_LQ_R + H.c.)\,,
\label{L-UV}
\eea
where, we assume $y_Q \gg y_\Psi$. After the spontaneous symmetry breaking, $\langle\Phi\rangle=f_a/\sqrt{2}$, the radial mode $\rho$ obtains a mass, $m_\rho\sim \sqrt{\lambda} f_a$ (where $\lambda$ is the quartic coupling in $V(\Phi)$), and the fermions acquire masses,
\beq
m_\Psi = \frac{y_\Psi f_a}{\sqrt{2}}, ~~~~~{\rm and}~~m_Q = \frac{y_Q f_a}{\sqrt{2}}\,.
\eeq
The Lagrangian involving DM mass becomes
\beq
\mathcal{L} \supset - m_\Psi  \overline{\Psi}_L\Psi_R e^{ia/f_a}.
\label{Psi-mass}
\eeq
The phase factor $e^{ia/f_a}$ in Eq.~\eqref{Psi-mass} can be removed via a field redefinition such as 
\beq
\Psi \to \Psi e^{i\gamma_5 a/(2f_a)}.
\label{field-redefinition}
\eeq
Such field redefinition in turn generates the standard ALP-DM interaction once we invoke Eq.~\eqref{field-redefinition} into the DM kinetic term in Eq.~\eqref{L-UV}, which yields
\beq
\mathcal{L}=-\frac{1}{2 f_a} \partial_\mu a \overline{\Psi} \gamma^\mu \gamma_5 \Psi,
\label{eq34}
\eeq
where the shift-symmetry of ALP is maintained. This is the standard derivative coupling expected for pseudo-Goldstone bosons.
Also, by expanding the $e^{ia/f_a}$ in Eq.~\eqref{Psi-mass}, one can get an alternate ALP-DM interaction term as
\beq
\mathcal{L}=-i\frac{m_\Psi}{f_a} a \overline{\Psi} \gamma_5 \Psi,
\eeq
 which is equivalent to Eq.~\eqref{eq34} at the leading order in $1/f_a$. A similar chiral rotation applied to the heavy fermions $Q$ generates effective ALP couplings to electroweak gauge bosons through axial anomaly. In other words, the Noether current corresponding to this $U(1)$ symmetry (whose spontaneous breaking originates the ALP), $J^\mu_{U(1)}$, receives an anomalous contribution at the quantum level due to the presence of heavy VLFs charged under the electroweak symmetry. Its divergence is given by the triangle anomaly
	\beq
	\partial_\mu J^\mu_{U(1)}=\frac{g^{\prime 2} E_B}{16\pi^2}B_{\mu\nu}\tilde{B}^{\mu\nu}+\frac{g^2 E_W}{16\pi^2}W_{\mu\nu}\tilde{W}^{\mu\nu}.
	\eeq
Here, $E_B$ and $E_W$ are the anomaly coefficients, given by \cite{Srednicki:1985xd,DiLuzio:2020wdo}
$$E_B=\sum_i X_i Y_i^2,~~~~E_W=\sum_i X_i T_2(R_i),$$
where $X_i,Y_i, T_2(R_i)$ denote the $U(1)$ charge, hypercharge, and $SU(2)$ index of the heavy fermions, respectively.

\noindent As the ALP couples derivatively to the $U(1)$ current as
$\mathcal{L}\supset \partial_\mu a J^\mu_{U(1)}/{f_a}$, applying integration by parts together with the anomalous divergence, one can easily obtain the following interactions between ALP and gauge bosons:
\beq
\mathcal{L}_{\rm eff}=\frac{a}{f_a}\left[\frac{g^{\prime 2} E_B}{16\pi^2}B_{\mu\nu}\tilde{B}^{\mu\nu}+\frac{g^2 E_W}{16\pi^2}W_{\mu\nu}\tilde{W}^{\mu\nu} \right],
\eeq
which we have effectively written in Eq. \ref{eq:ALP-SM} as
\beq
\mathcal{L}\supset \frac{1}{4}C_{aX}~ a ~X_{\mu\nu}\tilde{X}^{\mu\nu}~~~~(X=B,W),
\eeq
%In our effective parametrization in Eq.~\eqref{eq:ALP-eff}, these interactions are written as
with the coefficients defined as
\beq
C_{aB}=\frac{g^{\prime 2} E_B}{4\pi^2 f_a}\equiv \frac{\alpha_1}{\pi f_a}E_B,~~{\rm and}~~~C_{aW}=\frac{\alpha_2}{\pi f_a}E_W.
\eeq
Thus, the Wilson coefficients already include the gauge couplings and the one-loop suppression factor, satisfying $C_{\alpha i}\propto \alpha_i/(\pi f_a)$, such that, after EWSB, the physical ALP-photon coupling is given by $g_{a\gamma\gamma}\sim \alpha_{\rm EM}/{\pi f_a}$, reproducing the standard result obtained from the anomalous triangle diagram. The anomaly coefficient $E_B$ and $E_W$ depend on the specific $U(1)$ charges and electroweak quantum number of the heavy VLFs in the UV completion. Notably, here, we did not fix a particular charge assignment and instead adopt the natural benchmark $E_B\sim E_W\sim \mathcal{O}(1)$, corresponding to minimal VLF content with order-one charges.

\noindent On the other hand, for a perturbative UV completion, $y_Q, \lambda\lesssim 4\pi$, implying,
\beq
\Lambda_{\rm eff}\lesssim \mathcal{O}(4\pi f_a)\,.
\eeq
Hence, $4\pi f_a$ is regarded as a conservative upper bound rather than a strict cutoff. For the benchmark values considered in our analysis, $f_a \sim \mathcal{O}(200)$ GeV and anomaly coefficients of order unity naturally yield $g_{a\gamma\gamma}\sim \mathcal{O}(10^{-5})$ GeV$^{-1}$, which is sufficient to reproduce the observed relic abundance. At the same time $\Lambda_{\rm eff}\sim \mathcal{O}$(TeV) and the relevant energy scales satisfy $\sqrt{s},m_a,m_\Psi\ll \Lambda_{\rm eff}$. Therefore, the loop-induced ALP–photon coupling can be consistently realized within the regime of validity of the effective field theory. This setup, resembling the KSVZ model \cite{Kim:1979if,Shifman:1979if} in the context of QCD axion, therefore allows for ALP-mediated DM interactions and anomaly induced couplings relevant for astrophysical and cosmological probes, while maintaining consistence with flavor and collider bounds.

%Similarly, the heavy exotic quarks will also obtain a mass as $m_Q = y_Q f_a/\sqrt{2}$. However, as $Q$ fields are also charged under SM electroweak symmetry, a
%similar field redefinition on $Q$ such as in Eq.~\eqref{field-redefinition} generates effective ALP couplings to SM electroweak gauge bosons through the axial anomaly. Therefore, at energies well below $f_a$, the effective interactions of ALPs can be generated as written in Eq.~\eqref{eq:ALP-SM} where these exotic quarks appear as the intermediate states. On the other hand, ALP-SM fermion couplings do not appear in the scenario as SM fermion fields do not couple to $\Phi$ and carry no $U(1)$ charge. This setup, resembling the KSVZ model \cite{Kim:1979if,Shifman:1979if} in the context of QCD axion, allows for ALP-mediated DM interactions and anomaly induced couplings relevant for astrophysical and cosmological probes, while maintaining consistence with flavor and collider bounds.

%
%%%%%%%%%%%%%%%%%%%%%%%%%%%%%%%%%%%%%%%%%%%%%%%%%%%%%%%%%%%%%%
\section{Mono-photon signal at the HL-LHC}
	\label{sec:hl-lhc}
%%%%%%%%%%%%%%%%%%%%%%%%%%%%%%%%%%%%%%%%%%%%%%%%%%%%%%%%%%%%%%

\begin{figure}[htb!]
	$$
	\includegraphics[scale=0.385]{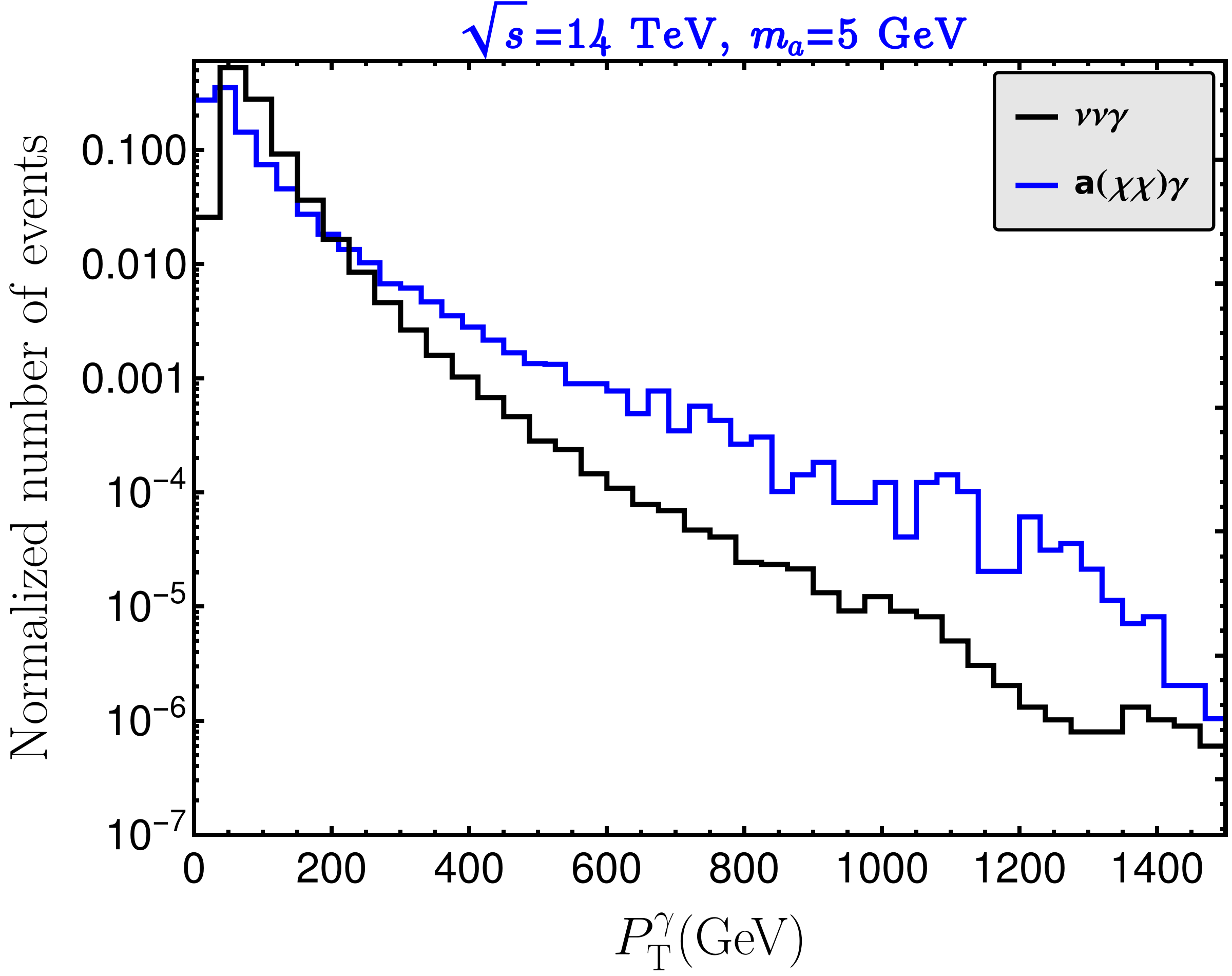}~~
	\includegraphics[scale=0.4]{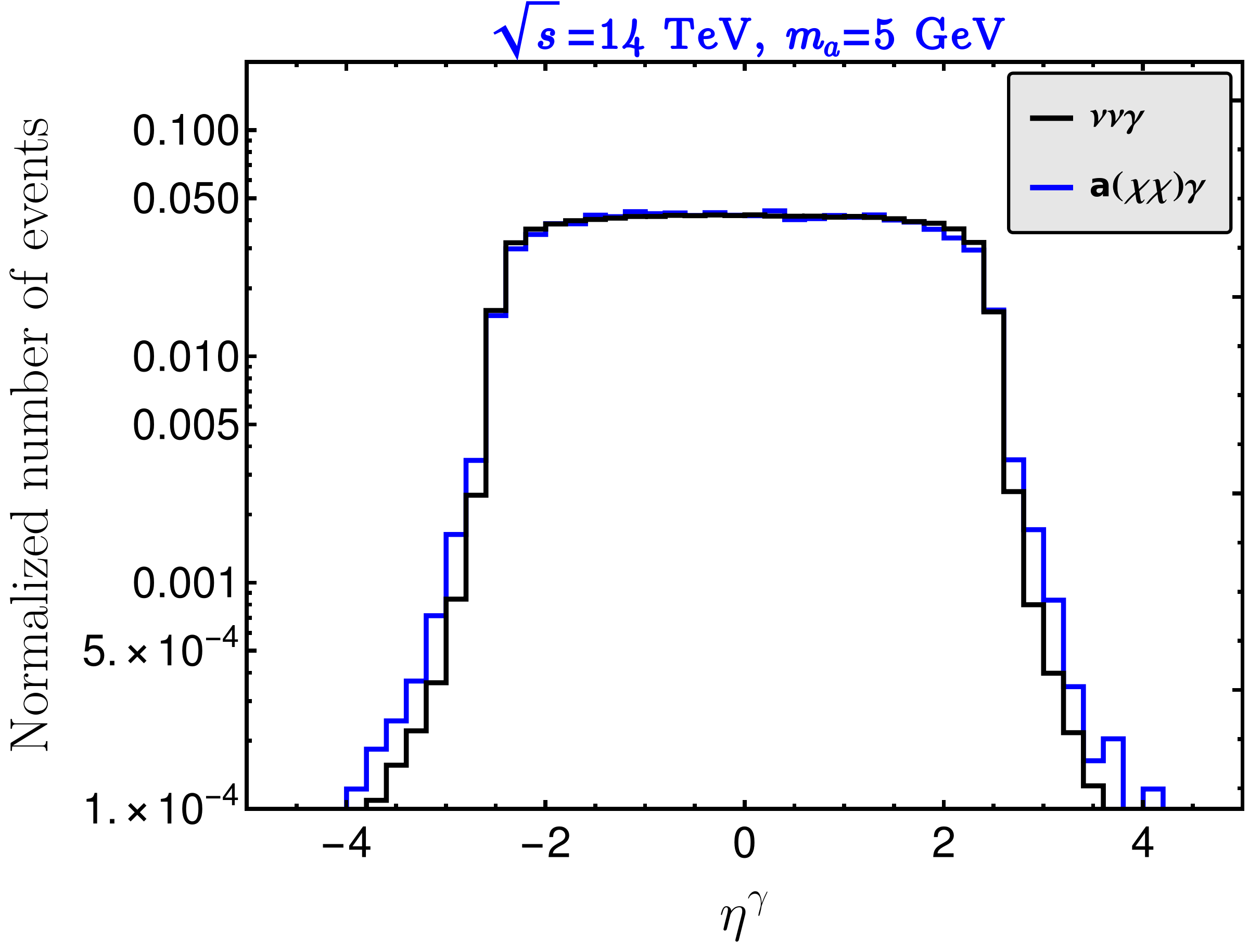}
	$$
	$$
	\includegraphics[scale=0.4]{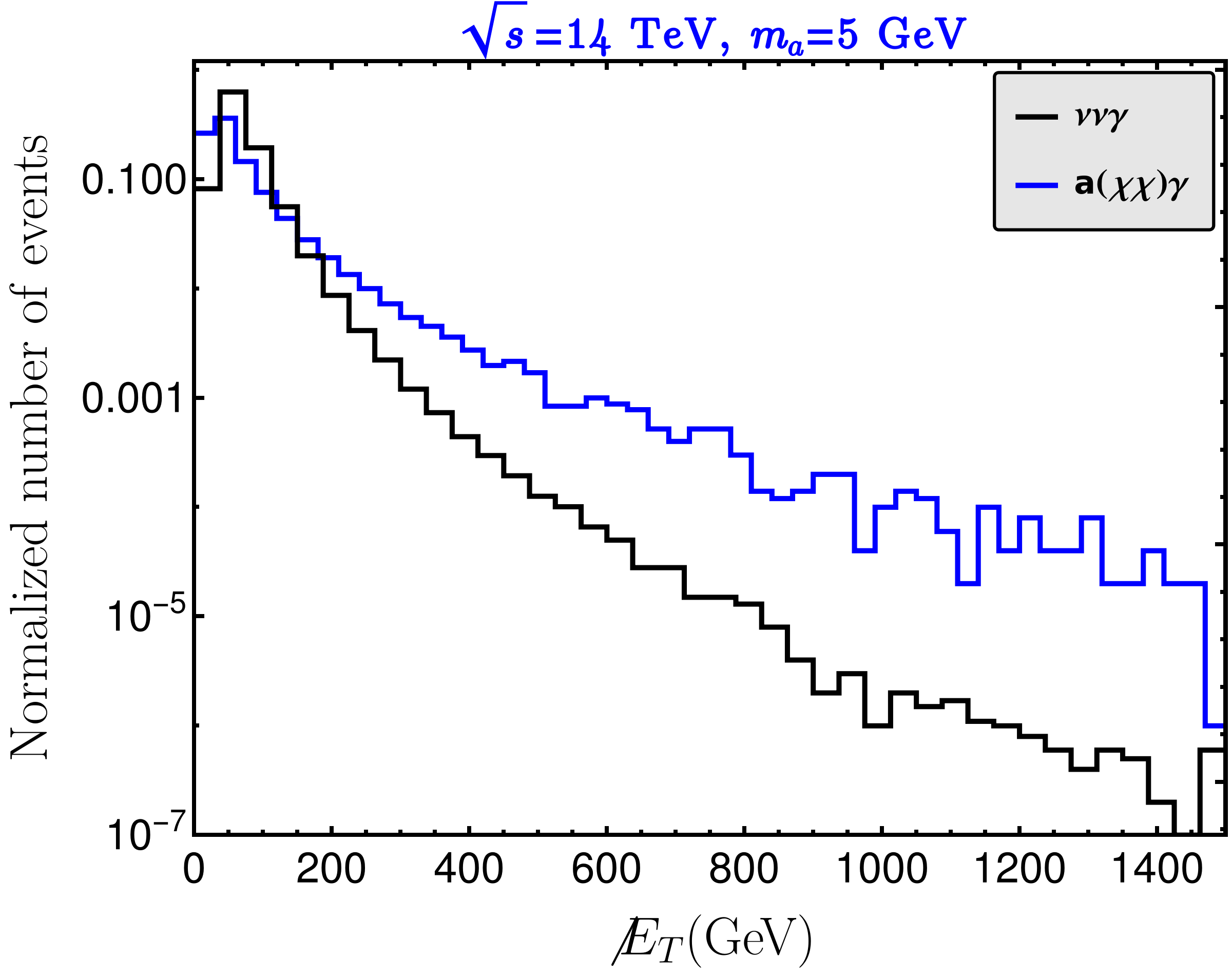}~~
	\includegraphics[scale=0.4]{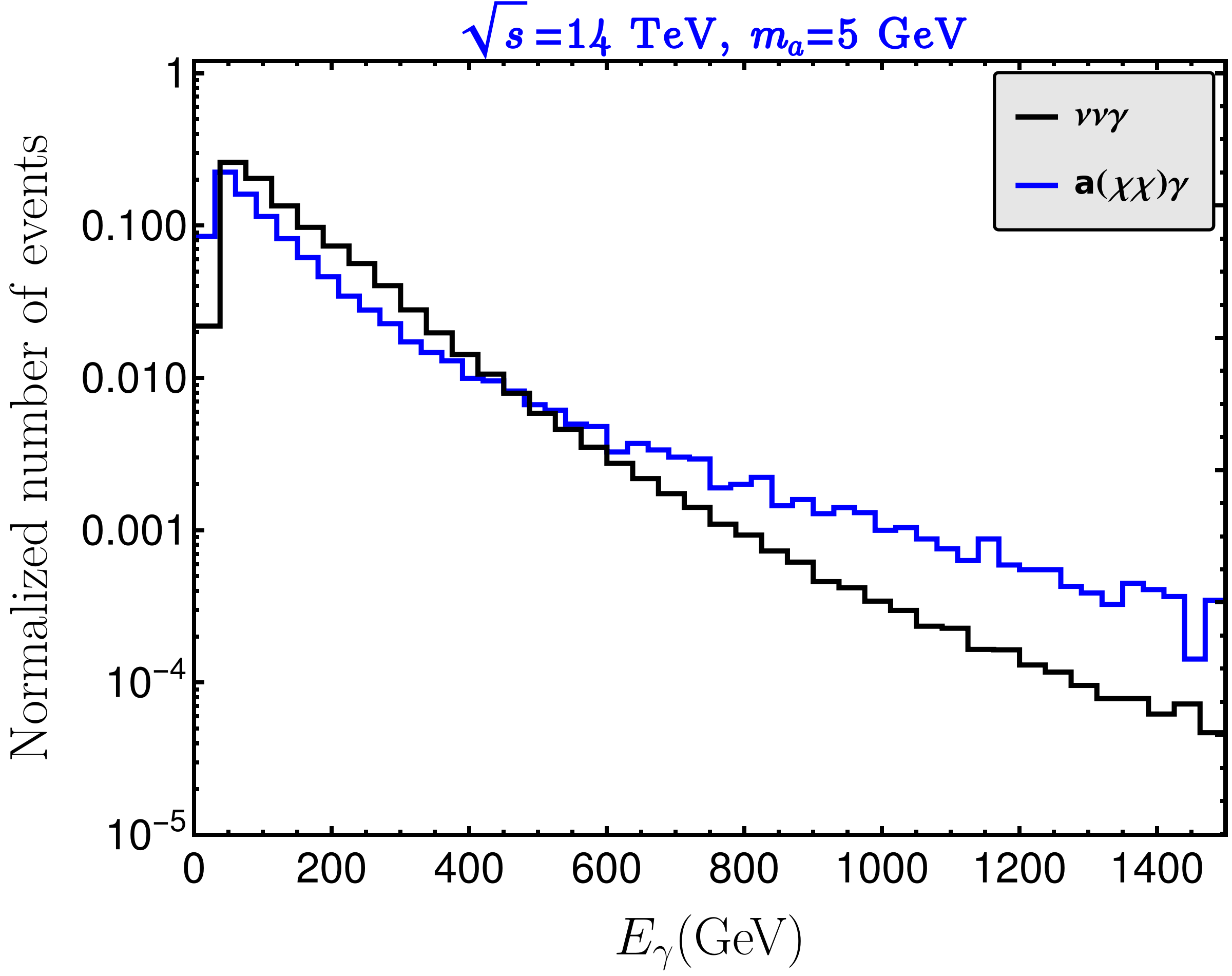}
	$$
	\caption{Normalized event distribution of different kinematic variables at the HL-LHC with $\sqrt{s}=14$ TeV and $m_a=5$ GeV. Top left: transverse momentum of photon($P^{\gamma}_T$), top right: pseudorapidity ($\eta_{\gamma}$), bottom left: MET ($\slashed{E}_{T}$), bottom right: photon energy ($E_{\gamma}$).}
	\label{fig:event.hl.lhc}
\end{figure}
We discuss here the potential of the High-Luminosity LHC (HL-LHC) to probe ALP-portal DM with $\sqrt{s}=14$ TeV and $\mathfrak{L}_{\text{int}}=3~\text{ab}^{-1}$ via hadronically quiet mono-photon plus missing energy signal. At the hadron colliders, as we do not have the knowledge of subprocess CM energy, we need to rely upon the transverse momentum of the visible particles to cook up missing transverse energy to be associated with DM particles that are produced; defined as follows; 
\begin{itemize}
\item {\bf Missing transverse energy ($\slashed{E}_T$):} $\slashed{E}_T$ arises from the imbalance in the total 
momentum measured in the plane perpendicular to the beam axis and can be calculated as
	\begin{equation}
		\slashed{E}_T = -\sqrt{\left(\sum p_x\right)^2 + \left(\sum p_y\right)^2}\,,
	\end{equation}
	where $p_x$ and $p_y$ denote the components of the momenta along the $x$ and $y$ directions respectively, summed over 
	all the visible particles registered at the detector.
	\end{itemize}

Here we adopt a simple cut based analysis technique, as this aims to be indicative than exhaustive. After applying the photon selection criteria, requiring $p_T^\gamma > 20~\text{GeV}$ and $|\eta^\gamma| < 3$, we find that the signal cross-section is $1.20~\text{fb}$, while the dominant SM background ($\nu \bar{\nu} \gamma$) yields a cross-section of $6.10~\text{pb}$. The normalized event distributions of different kinematical variables are shown in Fig.~\ref{fig:event.hl.lhc}. The distributions show 
that there is no potential distinction between the signal and dominant SM background at LHC. Roughly 2000 signal events are left after the basic selection cut whereas the background events are 18 millions. 
Once we employ a lower cut $\slashed{E}_T>1$ TeV, we have only 3 signal events left and the leftover 
background events are still huge $\sim$ 5000. Similarly no other kinematical variable shown in Fig.~\ref{fig:event.hl.lhc} 
yield a signal significance worth mentioning. Thus we infer that a although a dedicated machine learning analysis can possibly do better concerning HL-LHC, given that none of the variables have a promising segregation 
between the signal and the background, probing ALP-photon coupling at the HL-LHC seems difficult.

%%%%%%%%%%%%%%%%%%%%%%%%%%%%%%%%%%%%%%%%%%%%%
\section{Opposite sign muon signal}
\label{sec:osms}
%%%%%%%%%%%%%%%%%%%%%%%%%%%%%%%%%%%%%%%%%%%%%
%%%%%%%%%%%%%%%%%%%%%%%
\begin{figure}[htb!]
	\centering
	\begin{tikzpicture}
		\begin{feynman}
			\vertex (a) at (-1, 1.6) {\Large $\mu^-$};
			\vertex (b) at (-1, -1.6) {\Large $\mu^+$};
			\vertex (c) at (2, 1.6);
			\vertex (g1) at (2, 0);
			\vertex (g2) at (5, 0) {\Large $a$};
			\vertex (d) at (2, -1.6);
			\vertex (e) at (5, 1.6) {\Large $\mu^-$};
			\vertex (f) at (5, -1.6) {\Large $\mu^+$};
			
			\diagram* {
				(a) -- [fermion, ultra thick, arrow size=2pt] (c) -- [fermion, ultra thick, arrow size=2pt] (e),
				(b) -- [anti fermion, ultra thick, arrow size=2pt] (d) -- [anti fermion, ultra thick, arrow size=2pt] (f),
				(c) -- [boson, ultra thick, edge label=$\gamma$] (g1) -- [boson, ultra thick, edge label=$\gamma$] (d), 
				(g1) -- [scalar, ultra thick] (g2)
			};
			\vertex at (g1) [blob, minimum size=0.5cm, fill=gray!50] {};
		\end{feynman}
	\end{tikzpicture}
	\caption{Feynman graph for ALP production via photon fusion at the muon collider.}
	\label{fig:ph.fusion}
\end{figure}

A promising collider signature may arise from the opposite-sign muon (OSM) channel at the $\mu^+\mu^-$ collider, 
where ALP is produced in association with OSM pair via the photon fusion radiated from the initial-state muons, commonly referred to as \textit{vector boson fusion (VBF)}, as illustrated in Fig.~\ref{fig:ph.fusion}. The axion decaying to DM thus produces OSM + $\slashed{E}$ signal. The dominant irreducible SM background contributions originate from processes such as $W^+W^-$ and $\tau^+\tau^-$, where both of the gauge bosons and tau leptons decay into muons and neutrinos. Additional significant backgrounds arise from $\mu^+\mu^-Z$ production, with the $Z$ boson decaying invisibly to neutrinos, and from $\nu\bar{\nu}Z$ production, where the $Z$ boson decays into a $\mu^+\mu^-$ pair.

The generation of signal and background events, along with parton showering, hadronization, and detector simulation are carried out as described in the main text. For the OSM signal, we strictly select events containing exactly two muons, vetoing events with any photons, jets, or electrons in the final state. In the event generation stage, the phase space is restricted by requiring a minimum transverse momentum of $p_T^\mu > 10$~GeV for the muon, along with a pseudorapidity of $|\eta_{\mu}| \leq 3$. Along with $E_{\text{miss}}, \slashed{E}_T$, other key kinematic variables relevant to the OSM signal include
\begin{itemize}
	\item \textbf{Invariant di-muon mass} ($M_{\mu\mu}$): The invariant mass of the di-muon system is defined as  
	\begin{equation}
		M_{\mu\mu} = \sqrt{\left(p_{\mu^{+}} + p_{\mu^{-}}\right)^2}\,,
	\end{equation}  
	where \(p_{\mu^{+}}\) and \(p_{\mu^{-}}\) denote the four-momenta of the detected anti-muon and muon, respectively. Invariant mass distributions typically exhibit peaks at resonance positions and serve as a valuable discriminating variable, especially when the signal or background involves the resonant production of a heavy particle.
	
	\item {\bf Angular  separation between the muon Pair ($\Delta R_{\mu \mu}$):} The angular distance between the muon and anti-muon is defined in the $(\eta, \phi)$ plane as  
	\begin{equation}
		\Delta R_{\mu \mu} = \sqrt{\left(\Delta \eta_{\mu \mu}\right)^2 + \left(\Delta \phi_{\mu \mu}\right)^2}\,,
	\end{equation}  
	where $\Delta \phi_{\mu \mu}$ represents the difference in their azimuthal angles. This variable quantifies how well-separated the two particles are in the detector and is often used to characterize event topology.
\end{itemize}

\begin{figure}[htb!]
	$$
	\includegraphics[scale=0.385]{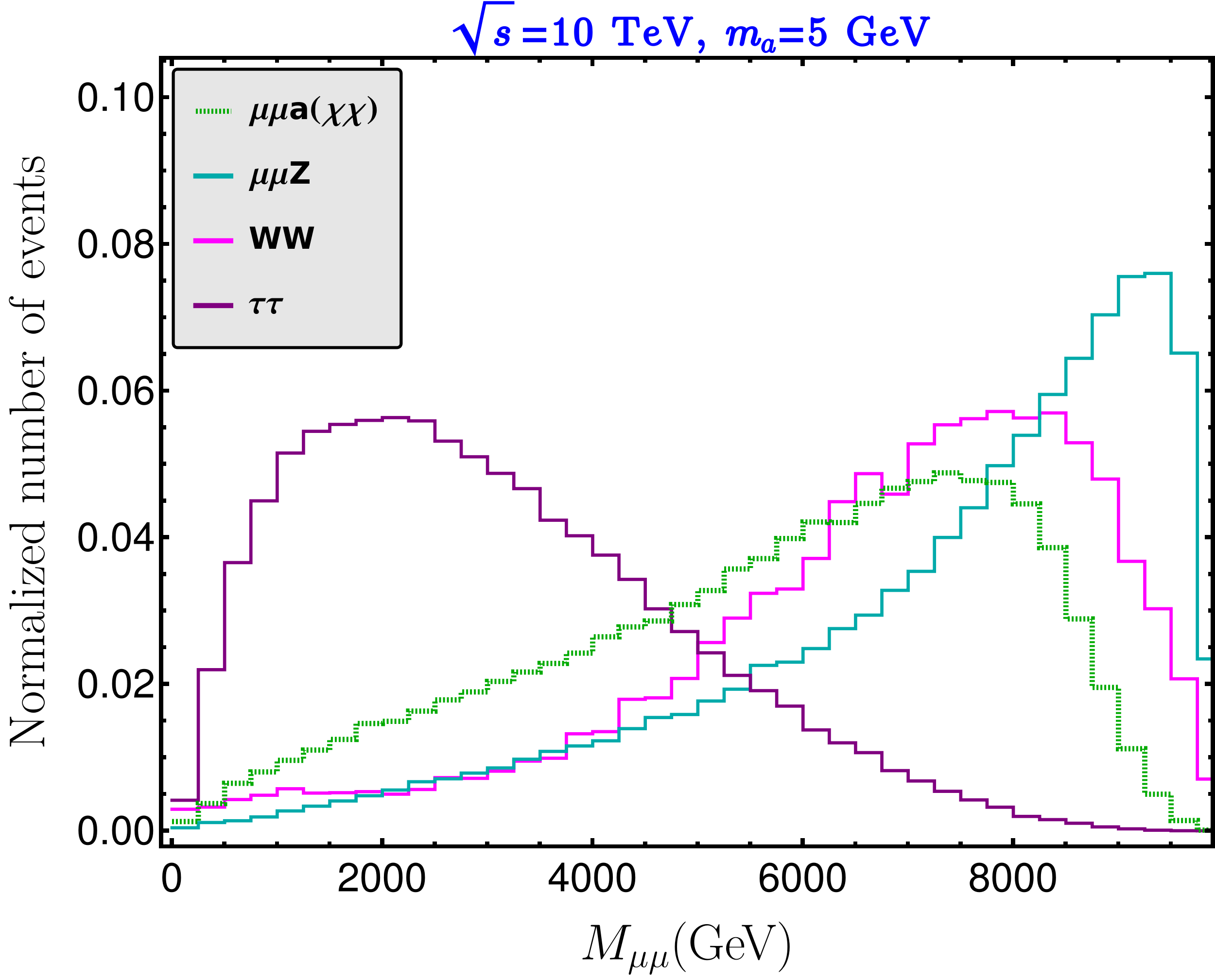}~~
	\includegraphics[scale=0.4]{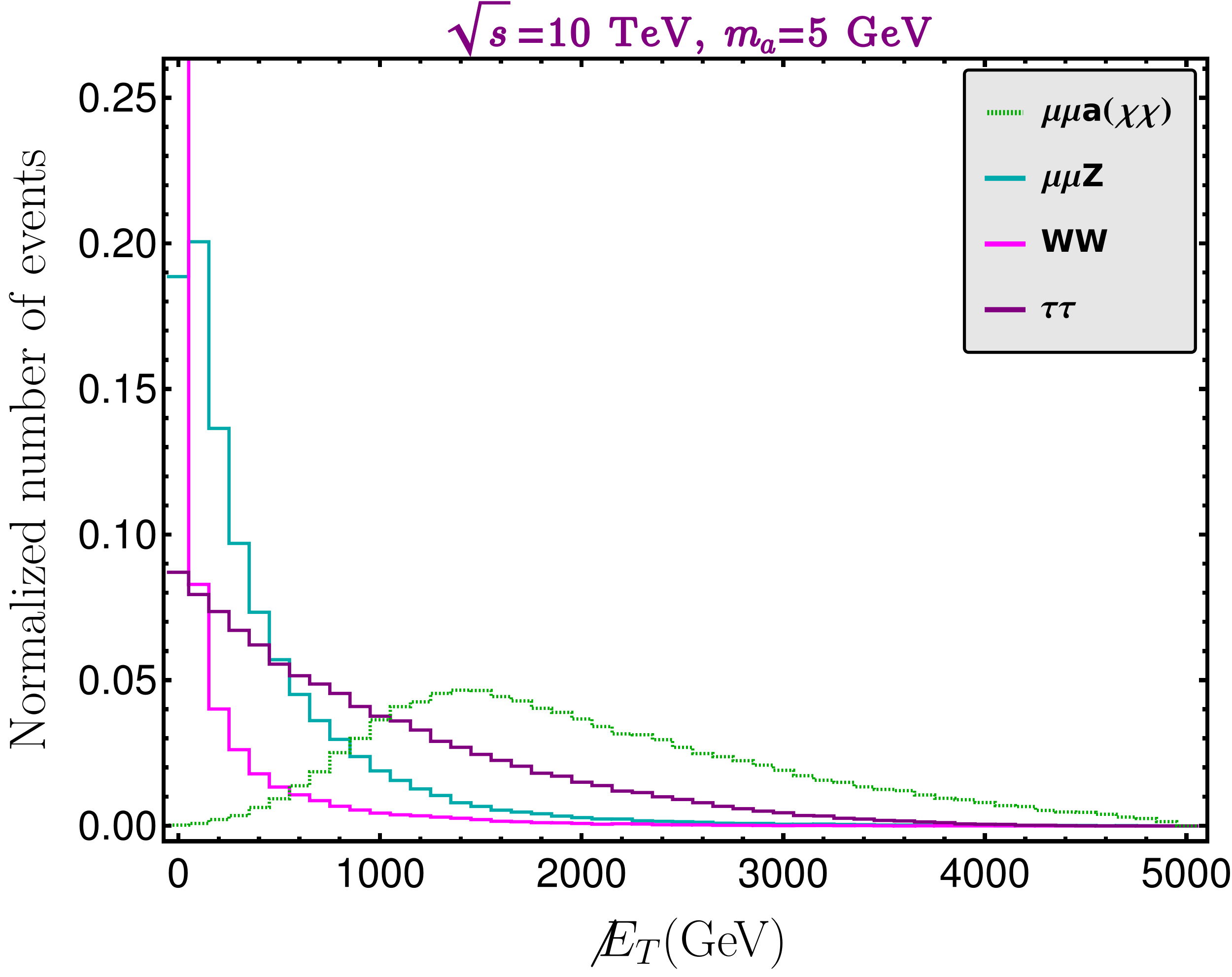}
	$$
	$$
	\includegraphics[scale=0.4]{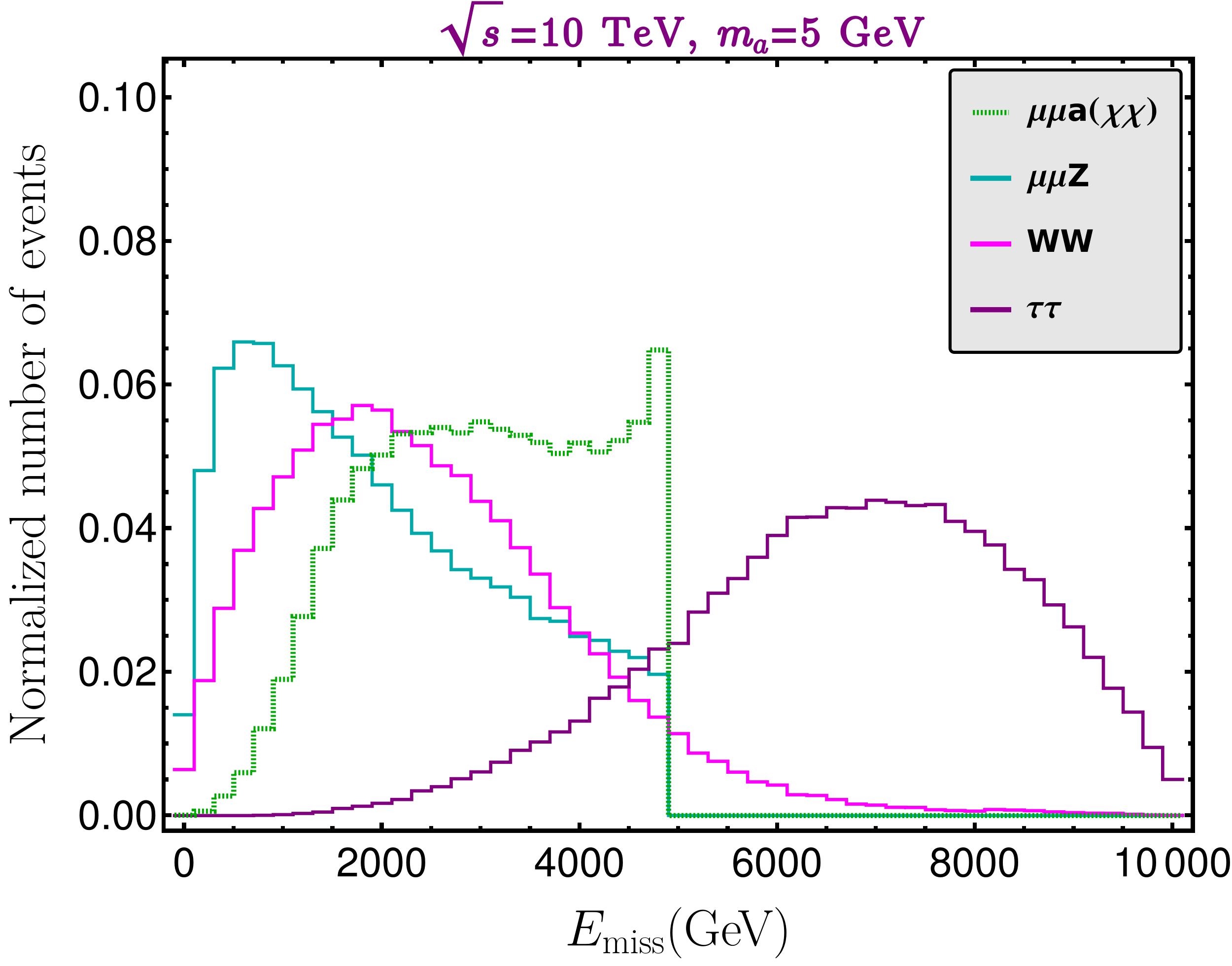}~~
	\includegraphics[scale=0.4]{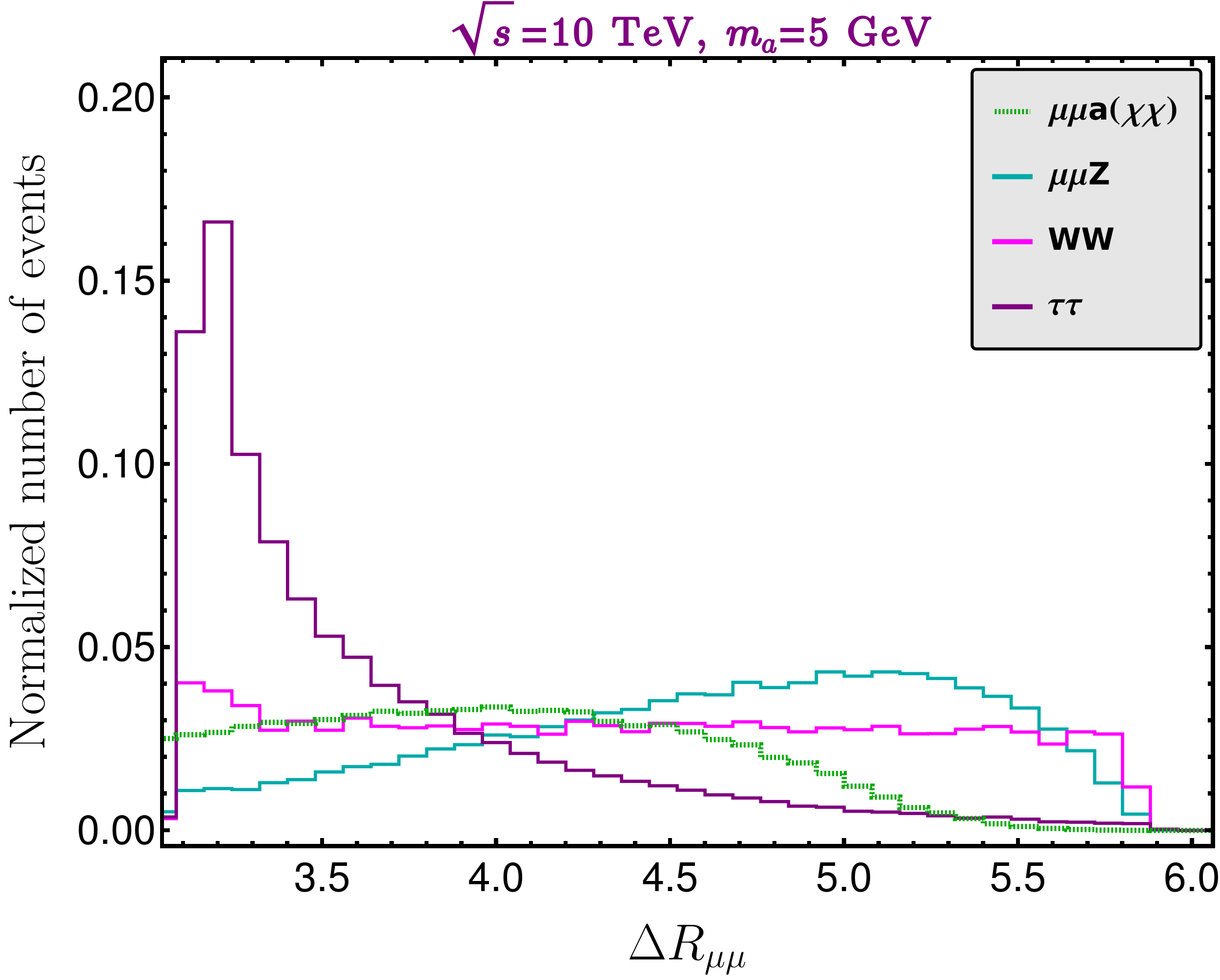}
	$$
	\caption{Normalized event distribution of kinematic variables at the muon collider with $\sqrt{s}=10$ TeV and $m_a=5$ GeV. Top left: invariant dimuon mass ($M_{\mu \mu}$), top right: missing transverse energy ($E_{\text{miss}}$), bottom left: missing  energy ($E_{\text{miss}}$), bottom right: angular separation ($\Delta R_{\mu \mu}$) .}
	\label{fig:event.dist.vbf}
\end{figure}

\begin{table}[h!]
	\centering
	\begin{tabular}{|c|c|ccc|c|c|}
		\hline
		\multirow{2}*{Cuts} & \multirow{2}*{Signal} & \multicolumn{3}{c|}{Backgrounds} & \multirow{1}*{Significance} \\
		\cline{3-5}
		&        & $\mu\mu Z$ & $WW$ & $\tau\tau$ &  ($\mathcal{Z}$)  \\
		\hline
		Basic cuts & 680 & 15400 & 6990 & 306  & 4.45 \\
		$\slashed{E}_T > 845$ GeV & 625 & 2088 & 250 & 131  & 11.23 \\
		$1.85~\text{TeV}<E_{\text{miss}} < 4.95$ TeV & 501 & 966 & 135 & 20  & 12.44 \\
		\hline
	\end{tabular}
	\caption{Cutflow for signal and SM background events for OSM final state at the muon collider with $\sqrt{s}$ = 10 TeV and $\mathfrak{L}_{\text{int}}$ = 10 $\rm{ab^{-1}}$. Note that the basic cuts include the selection criteria with invariant di-lepton mass cut.}
	\label{tab:cut.flow.muC}
\end{table}

The normalised signal and background event distributions are shown in Fig.~\ref{fig:event.dist.vbf}, where the distinction between signal and background is quite apparent in several variables. To suppress $\nu \nu Z$ background, a pre-selection cut of $|M_{\mu \mu}| > 100$ GeV is applied. This cut effectively eliminates contributions from the $Z$-pole region, thereby removing this background without harming the signal. Fig.~\ref{fig:event.dist.vbf} presents the normalized event distributions for both signal and background for kinematic variables discussed above. All the SM backgrounds, carry low $\slashed{E}_T$, while, the signal tends to have significantly higher $\slashed{E}_T$. This difference leads to a distinctive shape in the $\slashed{E}_T$ distribution, which allows for effective discrimination between the DM signal and the SM background, as evident in the top-right panel of Fig.~\ref{fig:event.dist.vbf}. Applying a cut of $\slashed{E}_T > 845$ GeV eliminates around 90\% of the SM background as a whole, while preserving nearly 90\% of the signal. A subsequent cut requiring $1.85~\text{TeV} < E_{\text{miss}} < 4.95~\text{TeV}$ further reduces the total SM background by approximately 55\%, while retaining around 80\% of the signal relative to the former cut. Although $\Delta R_{\mu\mu}$ could be employed as an additional variable to suppress the $\tau^+\tau^-$ background, it proves to be less effective in our case, as it does not lead to a significant improvement in the $\mathcal{Z}$. The signal and background events, along with  $\mathcal{Z}$ are presented in Table~\ref{tab:cut.flow.muC} after employing subsequent cuts considering $m_a=5$ GeV, $\sqrt{s}=10$ TeV and $\mathfrak{L}_{\text{int}}$ = 10 $\rm{ab^{-1}}$.  We note that the basic cuts the include selection cuts with the $|M_{\mu \mu}|>100$ GeV. Implementation of the $\slashed{E}_T$ cut improves the $\mathcal{Z}$ by approximately a factor of 2.5, while the $E_{\text{miss}}$ cut enhances $\mathcal{Z}$ by about 11\%.

%%%%%%%%%%%%%%%%%%%%%%%%%%%%%%%%%%%%%%
\section{Differential cross-section of $e^+e^- \to a\gamma$}
\label{sec:diff.cross}
%%%%%%%%%%%%%%%%%%%%%%%%%%%%%%%%%%%%%

Associate productions of ALPs $e^+e^- \to a\gamma$ has been mainly pursued in this paper via Feynman diagram shown in Fig.~\ref{fig:alp.prod}. We furnish the analytical formula for the differential production cross-section here. The amplitude of the process $e^+(p_2)e^-(p_1) \to \gamma (k_2) a(k_1)$ is given by
\begin{align}
	\mathcal{M}=\frac{e_0 g_{a \gamma \gamma}}{4s}\bar{v}(p_2)\gamma^{\mu}u(p_1)\epsilon_{\mu \nu \rho \sigma} k^\rho q^\sigma \epsilon^{* \nu}(k)\,,
\end{align}
with $q=(p_1+p_2)$ and $k=(q-k_1)$. $e_0$ denotes the $U(1)_{\text{EM}}$ charge and $\epsilon_{\mu \nu \rho \sigma}$ is the anti-symmetric tensor.
 The differential cross section for ALPs  produced in association with a $\gamma$ is given by 
\begin{align}
	\frac{d \sigma(e^+ e^-\to\gamma a)}{d\Omega} =
	& \frac{\alpha}{64 } g_{a\gamma \gamma}^2 \left(1 - \frac{m_a^2}{s}\right)^3 \left(1 + \cos^2 \theta \right), 
\label{eq:diff.cross.ALP}
\end{align}
where $\alpha=e_0^2/4 \pi$ is the fine structure constant and $\cos\theta$ is the scattering angle of photon in the CM frame.

%%%%%%%%%%%%%%%%%%%%%%%%%%%%%%
\bibliographystyle{JHEP}
\bibliography{ref}
%%%%%%%%%%%%%%%%%%%%%%%%%%%%%%
\end{document}